\documentclass[twocolumn,showpacs,unsortedaddress,superscriptaddress,showkeys,nofootinbib,preprintnumbers,letterpaper]{revtex4-1}
\usepackage{bm}
\usepackage{upgreek}
\usepackage{acronym}
\usepackage{amsmath}
\usepackage{amssymb}

\usepackage{graphicx}
\usepackage[ruled,vlined]{algorithm2e}

\usepackage[caption=false]{subfig}

\usepackage{siunitx}
\usepackage[usenames]{color}
\usepackage{comment}
\usepackage{xspace}
\usepackage[normalem]{ulem}
\usepackage[%
 setpagesize=false,%
 bookmarks=true,%
 bookmarksdepth=tocdepth,%
 bookmarksnumbered=true,%
 colorlinks=false,%
 pdftitle={},%
 pdfsubject={},%
 pdfauthor={},%
 pdfkeywords={}%
]{hyperref}

\newcommand*{\ee}{\mathrm{e}}

\newcommand{\diff}[1]{\,\mathrm{d}{#1}\,}
\newcommand*{\difff}{\mathrm{d}}

\newcommand{\average}[1]{\left\langle{#1}\right\rangle}

\newcommand{\Msun}{M_{\odot}}

\newcommand{\ho}{{H_0}}
\newcommand{\Om}{{\Omega_\mathrm{m}}}
\newcommand{\bRs}{{\bar{R}_\mathrm{s}}}
\newcommand{\bRn}{{\bar{R}_\mathrm{n}}}
\newcommand{\bRtot}{{\bar{R}_\mathrm{tot}}}
\newcommand{\etabar}{{\bar{\eta}}}
\newcommand{\vrho}{{\vec{\rho}}}
\newcommand{\vxisq}{{\vec{\xi^2}}}
\newcommand{\Ntot}{{N_\mathrm{tot}}}
\newcommand{\Ndet}{{N_\text{inst}}}

\newcommand{\pastro}{{p(\mathrm{astro})}}
\newcommand{\hsignal}{{\mathcal{H}_\text{s}}}
\newcommand{\hnoise}{{\mathcal{H}_\text{n}}}
\newcommand{\lc}{{\lambda_\mathrm{c}}}
\newcommand{\lm}{{\lambda_\mathrm{m}}}

\newcommand{\mchirpd}{{\mathcal{M}_\mathrm{c}}}
\newcommand{\mchirps}{{M_\mathrm{c}}}
\newcommand{\pmd}{{m_{1, \text{d}}}}
\newcommand{\smd}{{m_{2, \text{d}}}}
\newcommand{\pms}{{m_{1, \text{s}}}}
\newcommand{\sms}{{m_{2, \text{s}}}}
\newcommand{\rcomb}{{\rho_\mathrm{comb}}}

\newcommand{\ropt}{{\rho_\mathrm{opt}}}
\newcommand{\robs}{{\rho_\mathrm{obs}}}
\newcommand{\Xobs}{{X_\mathrm{obs}}}
\newcommand{\Xexp}{{X_\mathrm{exp}}}
\newcommand{\mG}{{\mathcal{G}}}
\newcommand{\dc}{{d_\mathrm{C}}}
\newcommand{\dl}{{d_\mathrm{L}}}
\newcommand{\Vc}{{V_\mathrm{c}}}

\newcommand{\gaussian}{{\mathcal{N}}}
\newcommand{\lr}{{\mathcal{L}}}
\newcommand{\rsth}{{x_\text{th}}}
\newcommand{\template}{{\bar{\theta}}}
\newcommand{\ptrue}{{p_\text{true}}}
\newcommand{\phat}{{\hat{p}}}
\newcommand{\frachf}{{\mathcal{F}_\text{HF}}}

\newcommand{\mdagpstime}{{866736411--875232014}}

\newcommand{\figref}[1]{Fig.~\ref{#1}}
\newcommand{\secref}[1]{Sec.~\ref{#1}}
\newcommand{\algref}[1]{Algorithm~\ref{#1}}

\DeclareSIUnit{\annum}{a}
\DeclareSIUnit{\AU}{AU}
\DeclareSIUnit{\parsec}{pc}

\begin{document}
\preprint{RESCEU-29/25}


\title{Estimation of the Hubble parameter from unedited compact object merger catalogues}

\author{Reiko Harada}
\email{harada.reiko@resceu.s.u-tokyo.ac.jp}
\affiliation{Research Center for the Early Universe (RESCEU), Graduate School of Science, The University of Tokyo, Tokyo 113-0033, Japan}
\affiliation{Department of Physics, Graduate School of Science, The University of Tokyo, Tokyo 113-0033, Japan}
\author{Heather Fong}
\affiliation{University of British Columbia, Vancouver Campus, 2329 West Mall, Vancouver, British Columbia V6T 1Z4, Canada}
\author{Kipp Cannon}
\email{kipp@resceu.s.u-tokyo.ac.jp}
\affiliation{Research Center for the Early Universe (RESCEU), Graduate School of Science, The University of Tokyo, Tokyo 113-0033, Japan}

\begin{abstract}
In recent years, constraints on the Hubble parameter using multiple dark sirens have been made, relying on a galaxy catalogue, correlations between the mass and redshift distributions, or both.
	Those studies have typically used only significant gravitational wave candidates.
	In this work, we present a framework for cosmological inference that bypasses per-candidate parameter estimation, uses only detection-level information.
	This allows the population inference from a candidate list produced directly by a search pipeline, without additional selection cuts.
	Our method is particularly suited to extracting information from marginal candidates, which are essential for probing the distant universe.

\end{abstract}


\maketitle
\acrodef{CBC}{compact binary coalescence}
\acrodefplural{CBC}[CBCs]{compact binary coalescences}

\acrodef{SNR}{signal-to-noise ratio}
\acrodefplural{SNR}[SNRs]{signal-to-noise ratios}

\acrodef{PDF}{probability density function}
\acrodefplural{PDF}[PDFs]{probability density functions}

\acrodef{GW}{gravitational wave}
\acrodefplural{GW}[GWs]{gravitational waves}

\acrodef{FAR}{false-alarm rate}

\acrodef{IFAR}{inverse false-alarm rate}

\acrodef{LR}{likelihood ratio}
\acrodefplural{LR}[LRs]{likelihood ratios}

\acrodef{MDA}{mock data analysis}
\acrodefplural{MDA}[MDAs]{mock data analyses}

\acrodef{LVK}{LIGO--Virgo--KAGRA}

\acrodef{BBH}{binary black hole}
\acrodefplural{BBH}[BBHs]{binary black holes}

\acrodef{BNS}{binary neutron star}
\acrodefplural{BNS}[BNSs]{binary neutron stars}

\acrodef{EM}{electromagnetic}
\acrodef{PE}{parameter estimation}
\acrodef{KS}{Kolmogorov-Sminov}

\acrodef{p-p}{probability-probability}

\acrodef{RESCEU}{the Research Center for the Early Universe}

\section{Introduction}\label{sec:introduction}

The idea of using \ac{GW} to measure the cosmological parameters was first proposed by Schutz (1986), as the ``standard siren'' idea \cite{Schutz_1986}.
This vision became reality decades later, with the LIGO/Virgo detection of the \ac{BNS} merger GW170817, where the \ac{EM} counterpart enabled host-galaxy identification and thus a redshift, which combined with the GW-inferred luminosity distance to yield a direct $\ho$ measurement \cite{GW170817_first_standard_siren}.
GW170817 remains the only bright siren (with confirmed \ac{EM} counterpart) to date; most \ac{GW} detections are \ac{BBH} mergers that lack observable \ac{EM} signatures \cite{GWTC-1, GWTC-2, GWTC-2.1, GWTC-3, GWTC-4.0}.
Consequently, recent research has turned to ``dark siren'' methods: statistically inferring cosmological parameters from \ac{GW} candidates without \ac{EM} counterparts \cite{Pozzo_2012}.
These methods typically utilize external information such as galaxy catalogues \cite{GW170817_wo_em, Soares_santos_2019, Gray_2020, gwtc-3_cosmo, gwtc-4_cosmo} or any other assumed source redshift distributions \cite{Petiteau_2011, icarogw, gwtc-3_cosmo, gwtc-4_cosmo, Beirnaert_2025} in a hierarchical Bayesian framework.
In essence, cosmological parameters including $\ho$ are treated as a population parameters governing the spatial distribution of \ac{CBC} candidates, and constraints on its value are obtained by comparing the observed distribution of luminosity distances (or equivalently, \ac{SNR}) with the assumed redshift distribution.
A key feature of these analyses is that observed (detector-frame) masses are redshifted, entangling the mass distribution with cosmological parameters \cite{Mukherjee_2022, Yu_2022}.
This coupling both creates opportunities (e.g., ``spectral sirens'') \cite{Taylor_2012} and highlights the need to jointly infer cosmology and population parameters to avoid bias \cite{gwtc-3_cosmo, gwtc-4_cosmo}.
This is how cosmology has become an integral part of \ac{GW} population analysis---recent works incorporate $H_0$ into population models and estimate it jointly with astrophysical parameters.

Despite the progress, most \ac{GW} cosmology (population) studies so far have been limited to the most significant candidates.
For example, the GWTC-3 catalogue (through the end of LIGO/Virgo's third observing run) contained 90 \ac{CBC} candidates, yet cosmological inferences were performed using 47 significant candidates, with a network matched filter \ac{SNR}$>11$ and an \ac{IFAR} higher than 4 years \cite{gwtc-3_cosmo}.
Also, the latest \ac{LVK} analysis (GWTC-4.0, covering up to the first part of the fourth observing run) used 142 out of 218 total candidates to constrain $H_0$ and population properties, and the candidates are selected with the lowest \ac{FAR} among all search pipelines, ensuring all candidates have \ac{IFAR} higher than 4 years \cite{gwtc-4_cosmo}.

We consider that there are two main reasons for this reliance on a high significance threshold.
First, conventional \ac{GW} hierarchical inference requires a posterior samples in the \ac{CBC} parameter space for each candidate, obtained via full \ac{PE} \cite{Gaebel_2019, Gray_2020, Vitale_2021, icarogw, gwtc-3_pop, Gray_2023, gwtc-4_pop}.
This is a computationally expensive process and is thus usually conducted only for candidates with a high significance.
While recent advances (e.g. neural network surrogate models \cite{Gabbard_2021, DINGO, Bhardwaj_2023, AMPLFI, Aframe}) have enabled low-latency generation of \ac{PE} posteriors, applying \ac{PE} to every low-significance candidate is not yet standard.
Second, if one lowers the threshold to include marginal candidates, one inevitably introduces false alarms (noise-originating candidates) into the catalogue.
Analyzing a mixture of real signals and noise-originating candidates demands a multi-population modeling approach: one must simultaneously account for the astrophysical population of \acp{CBC} and the background ``population'' of noise candidates \cite{FGMC, Gaebel_2019, Shanika_2020, Roulet_2020, Vitale_2021, Heinzel_2023, Mehta_2025}.
However, implementing a rigorous joint inference with an unknown noise population is challenging---the distribution of noise-originating candidates in the \ac{CBC} parameter space is not well understood, and any mis-specification can bias the inferences.
Although some previous studies have proposed to indirectly incorporate the background population model of the search pipelines thorough the use of $\pastro$ \cite{Shanika_2020, Roulet_2020}, many population analyses to date have tended to avoid these complications by simply discarding lower-significance candidates.

In this paper, we propose a new framework for \ac{GW} population inference that is capable of incorporating candidates across a wide range of statistical significance, working directly with detection-level information.
Our approach operates on a \ac{CBC} catalogue produced by a \ac{GW} search pipeline, GstLAL \cite{Messick_2017}, and requires, for each \ac{CBC} candidate, only its detection statistic, together with the signal and noise models adopted in the search to compute significance (e.g. the models underlying \ac{FAR} \cite{Cannon_2013} and $\pastro$ \cite{Guglielmetti_2009, FGMC, BBHs_in_O1, rate_inferred_around_GW150914, Lynch_2018, Gaebel_2019, Kapadia_2020, Ray_2023} calculations).
Conventional hierarchical analyses generally rely on large-scale injection campaigns to estimate the detection efficiency and correct for selection biases \cite{Mandel_2019, Vitale_2021, Essick_2025}.
In contrast, since our dataset consists of detection statistics that already reflect the search threshold, our framework inherently incorporates the selection imposed by the search, without requiring explicit injection-based sensitivity estimates.

We develop this framework and demonstrate it as a proof of concept for cosmology.
In particular, we focus on estimating the Hubble constant from a mock catalogue of \acp{CBC}, showing how our approach can recover $\ho$ from the observational dataset.
We emphasize that the primary goal of this work is to introduce and validate the methodology, rather than to present scientific results of cosmological measurement.
Although our demonstration focuses on $\ho$, it can be extended to jointly infer $H_0$ alongside other population parameters (e.g. the merger rate evolution or mass distribution parameters).
It should be noted, however, that GstLAL's signal and noise models, as the \acp{PDF} of the detection statistic, are not known in analytic form; they are constructed numerically through importance sampling \cite{cannon2015likelihoodratiorankingstatisticcompact}.
Consequently, if the number of samples is insufficient or the resulting \acp{PDF} are not well converged, the residual fluctuations in the detection-statistic \acp{PDF} may exhibit dependence on the population parameters, which in turn can bias the population inference.
This issue, revealed through our \acp{MDA}, represents one of the key limitations of our current framework and points toward important directions for future improvement.

This paper is organized as follows.
In \secref{sec:method}, we describe our framework for population inference, only using detection level information.
\secref{sec:mda} describes a mock data analysis that serves as a proof-of-concept demonstration, including both the setup and the results.
\secref{sec:conclusion} summarizes our conclusions.

\section{Framework}\label{sec:method}
\subsection{Basics}\label{subsec:framework_basics}
The main goal here is to calculate a posterior distribution of the population parameters $p(\lambda|D, \Delta)$, where $D$ is our observational dataset, $\Delta$ denotes our detection criteria and observational limitation made by instrument's sensitivity and search algorithm.
Here, we define $\lambda$ as the set of all parameters that determine the shape of the \ac{CBC} population distribution, including the parameter describing the mass distributions of \acp{CBC}, $\lm$, and the cosmological parameters, $\lc$.
Heinzel et al.~(2023) introduced a framework that explicitly incorporates the glitch population parameters \cite{Heinzel_2023}.
In contrast, we do not aim to infer the terrestrial noise population jointly with \ac{CBC} population; instead, we treat the GstLAL noise model as a fixed, pre-characterized representation of the background population.
To be confirmed later, introducing $\Delta$ in our posterior is essential to account for selection effects.

In our framework, the dataset is interpreted not as raw strain data around each \ac{GW} candidate, but as the set of detection statistics assigned by a search pipeline to each \ac{GW} candidate.
Then, the dataset, $D$, is a \ac{CBC} catalogue of total size $\Ntot$, each of whose entries provides a detection statistic $x$, and our detection criteria, $\Delta$, is that the value of detection statistic is equal or above the threshold value: $x\geq \rsth$.
This approach eliminates the need for parameter estimation of individual \ac{GW} candidates and for large-scale injection campaigns to evaluate selection effects in catalogue construction.
As a result, it enables population analyses of \ac{GW} sources to be carries out at substantially lower computational cost, while incorporating even low-significance candidates.
Furthermore, since our observational dataset generally contains a mixture of candidates originating from \acp{CBC} and those arising from terrestrial noise, following the treatment introduced by Farr et al.~(2015) \cite{FGMC}, we account for the expected numbers, $\bRs$ and $\bRn$, of signal-originating and noise-originating candidates that satisfy the detection criteria $\Delta$ and occur during the observing period:
\begin{align}
	&p(\lambda|D, \Delta) = \int_0^{\infty}\diff{\bRs}\int_0^{\infty}\diff{\bRn}p(\lambda, \bRs, \bRn|D, \Delta),\\
	&=\int_0^{\infty}\diff{\bRtot}\int_0^1\diff{\etabar}p(\lambda, \bRtot, \etabar|D, \Delta),
	\label{eq:introduce_rate}
	\\
	&\propto\int_0^{\infty}\diff{\bRtot}\int_0^1\diff{\etabar}p(D|\lambda, \bRtot, \etabar, \Delta)\pi(\lambda, \bRtot, \etabar|\Delta),
	\label{eq:bayes_theorem}
\end{align}
where
\begin{equation}
	\bRtot = \bRs + \bRn
\end{equation}
is the expected total number of candidates, and
\begin{equation}
	\etabar = \frac{\bRs}{\bRs + \bRn}
\end{equation}
is the expected fraction of signal-originating candidates among the total.
This quantity is basically identical to the one denoted by $\bar{\eta}$ in \cite{Heinzel_2023}.
\eqref{eq:bayes_theorem} is derived from the Bayes' theorem.
The first term in \eqref{eq:bayes_theorem}, $p(D|\lambda, \bRtot, \etabar, \Delta)$, is the probability of obtaining the dataset given the hypotheses (the source population described by $\{\lambda, \bRtot, \etabar\}$, and the data acquisition process described by $\Delta$), which is called the hierarchical likelihood.
The second term in \eqref{eq:bayes_theorem}, $\pi(\lambda, \bRtot, \etabar|\Delta)$, is so called the prior probability and depends on our state of knowledge.
Although $\bRtot$ and $\etabar$ are, in principle, related to the population parameters $\lambda$ and our detection criteria $\Delta$, thorough the underlying source distribution, we treat them as free parameters in our inference.
Since their dependence on $\lambda$ should ultimately be refracted through the data, we do not model it explicitly here.
As a practical simplification, we assume a factorized prior of the following form:
\begin{equation}
	\pi(\lambda, \bRtot, \etabar|\Delta) = \pi(\lambda)\pi(\bRtot)\pi(\etabar).
	\label{eq:simplified_prior}
\end{equation}
In our \acp{MDA} in \secref{sec:mda}, we take $\pi(\bRtot)$ to be log-uniform over $[0, \infty)$ , as commonly done, and $\pi(\lambda)$ to be uniform within the prior range of $\ho$.
For the prior on $\etabar$, we consider two choices.
The first is a uniform over $[0, 1]$.
The second is a delta-function prior,
\begin{equation}
	\pi(\etabar) = \delta(\etabar - \etabar^\prime),
	\label{eq:dirac_delta_prior}
\end{equation}
where $\etabar^\prime$ denotes the point estimate of $\etabar$ obtained using the method described in \secref{subsec:fraction_estimate} based on $\pastro$.
Here, $\delta$ represents the Dirac delta function.

While our basic framework does not rely on a specific choice of detection statistic, $x$, the demonstration (\acp{MDA}) presented in \secref{sec:mda} uses the log-\ac{LR} ranking statistic, $\ln\lr$,  employed by GstLAL \cite{cannon2015likelihoodratiorankingstatisticcompact, Messick_2017, Tsukada_2023, gstlal}, and parts of the following discussion are written with this choice in mind.

\subsection{GstLAL likelihood-ratio ranking statistic}\label{sec:gstlal_lr}
The GstLAL \ac{LR} ranking statistic takes the form of
\begin{equation}
	\lr = \frac{p(\Theta|\hsignal)}{p(\Theta|\hnoise)}
	 = \frac{p(\vec{O}, \vrho, \vxisq, \vec{t}, \vec{\phi}, \bar{\theta}|\hsignal)}{p(\vec{O}, \vrho, \vxisq, \vec{t}, \vec{\phi}, \bar{\theta}|\hnoise)}.
	\label{eq:gstlal_LR}
\end{equation}
The numerator and denominator are the likelihoods of observing an $n$-dimensional set of parameters, $\Theta = \{\vec{O}, \vrho, \vxisq, \vec{t}, \vec{\phi}, \bar{\theta}\}$, under the signal and noise hypothesis, respectively.
This $n$-dimensional set includes the set of instruments involved in the observation of the candidate $\vec{O}$, the vectorized \ac{SNR} $\vrho$, the vectorized signal-based-veto parameter $\vxisq$, the vectorized arrival time $\vec{t}$, the vectorized coalescence phase $\vec{\phi}$, and the best match template parameters $\bar{\theta}$.
GstLAL provides models for both the numerator and denominator, and can evaluate them for the observed parameters \cite{cannon2015likelihoodratiorankingstatisticcompact, Tsukada_2023, gstlal}.
The evaluation relies not only on prior assumptions but also on empirical information, such as the time series of instrument horizon distances and instrumental noise properties derived from observational data products.
We refer to these models as the GstLAL signal model and the GstLAL noise model, respectively.

\begin{table*}
	\caption{Glossaries.}
	\begin{tabular}{cll}
		\hline\hline
		\multicolumn{2}{c}{$D=\{\{x\}, \Ntot\}$}& Observational dataset. \\
		& $x$ & Detection statistic of individual candidate.\\
		& $\Ntot$ & The total number of candidates.\\
		\hline
		$\Delta$ &\multicolumn{2}{l}{Detection criteria and observational limitation made by instruments' sensitivity and search algorithm.}\\
		\hline
		$\lc$ & \multicolumn{2}{l}{Cosmological parameters including $\ho$, $\Omega_\mathrm{m}$, \textit{etc.}}\\
		 &$\ho$& The Hubble constant.\\
		 &$\Om$& The fractional matter density.\\
		$\lm$ & \multicolumn{2}{l}{Parameters describing the mass distribution.}\\
		$\lambda$ & \multicolumn{2}{l}{Parameters describing source population in general.}\\
		\hline
		$\bRs$ & \multicolumn{2}{l}{The expected number of astrophysical candidates that satisfy $\Delta$.}\\
		$\bRn$ & \multicolumn{2}{l}{The expected number of background candidates that satisfy $\Delta$.}\\
		$\bRtot$ & \multicolumn{2}{l}{The expected total number of candidates.}\\
		$\etabar$ & \multicolumn{2}{l}{The expected fraction of signals (astrophysical candidates) among the total number of candidates.}\\
		\hline
		$\ln\lr$& \multicolumn{2}{l}{Log-\ac{LR} ranking statistic described in \eqref{eq:gstlal_LR} which is employed as GstLAL's detection statistic.} \\
		\multicolumn{2}{c}{$\Theta = \{\vec{O}, \vrho, \vxisq, \vec{t}, \vec{\phi}, \bar{\theta}\}$} & Set of GstLAL's ranking statistic variables describing a candidate.\\
		& $\vec{O}$ & Set of instruments involved in the observation of a candidate.\\
		& $\vrho$ & Vectorized \ac{SNR} (Set of \acp{SNR} observed by each instrument $\vec{O}$: $\vrho = \{\robs_{,O_1}, \robs_{,O_2}, \cdots\}$).\\
		& $\vxisq$ & Vectorized signals-based-veto parameter.\\
		& $\vec{t}$ & Vectorized arrival time.\\
		& $\vec{\phi}$ & Vectorized coalescence phase.\\
		& $\bar{\theta}$ & Best match template parameters.\\
		\hline
		$\hsignal$ & \multicolumn{2}{l}{A hypothesis that the candidate has an astrophysical origin.}\\
		$\hnoise$ & \multicolumn{2}{l}{A hypothesis that the candidate has a background origin.}\\
		\hline
		$z$ & \multicolumn{2}{l}{Redshift.}\\
		$\theta$ & \multicolumn{2}{l}{Set of intrinsic parameters describing a \ac{CBC} source (masses, spins, etc.).}\\
		$\robs$ & \multicolumn{2}{l}{\ac{SNR} observed by a instrument.}\\
		$\ropt$ & \multicolumn{2}{l}{Expected \ac{SNR} of a \ac{GW} signal from a \ac{CBC} which is optimally oriented and located to maximize the strain amplitude}\\
		 & \multicolumn{2}{l}{at a given instrument.}\\
		$\rcomb$ & \multicolumn{2}{l}{Combined \ac{SNR} of a instrument network.}\\
		\hline\hline
	\end{tabular}
	\label{table:glossaries}
\end{table*}

\subsection{Hierarchical likelihood}
In order to perform Bayesian parameter estimation described in \eqref{eq:bayes_theorem}, we need to model the hierarchical likelihood, $p(D| \lambda, \bRtot, \etabar, \Delta)$.
We will start from the following factorization:
\begin{align}
	&p(D| \lambda, \bRtot, \etabar, \Delta)
	=p(\{x\}, \Ntot|\lambda, \bRtot, \etabar, \Delta)\label{eq:derive1}\\
	&=p(\{x\}| \Ntot, \lambda, \bRtot, \etabar, \Delta)
	p(\Ntot|\lambda, \bRtot, \etabar, \Delta).\label{eq:derive2}
\end{align}
Assuming a Poisson process, the second term in \eqref{eq:derive2} depends on $\lambda$, $\etabar$ and $\Delta$ only thorough $\bRtot = \bRtot(\lambda, \etabar, \Delta)$; that is,
\begin{equation}
	p(\Ntot|\lambda, \bRtot, \etabar, \Delta) = p(\Ntot| \bRtot)
	=\frac{\bRtot^{\Ntot}\ee^{-\bRtot}}{\Ntot !}.
	\label{eq:Poisson}
\end{equation}
Assuming further that the detector noise has an autocorrelation length much smaller than the typical interval at which candidates are identified by the pipeline, and using $\hsignal$ and $\hnoise$ to denote the signal and the noise hypothesis (see Table \ref{table:glossaries}), the first term in \eqref{eq:derive2} can be written as:
\begin{align}
	&p(\{x\}| \Ntot, \lambda, \bRtot, \etabar, \Delta)=\prod^{\Ntot}_{i=1}p(x_i|\lambda, \bRtot, \etabar, \Delta)\label{eq:derive4}\\
	&=\prod^{\Ntot}_{i=1}
	\left[p(x_i|\hsignal, \lambda, \bRtot, \etabar, \Delta) p(\hsignal| \lambda, \bRtot, \etabar, \Delta)\right.\nonumber\\
	&\left. \hspace{7mm}+ p(x_i|\hnoise, \lambda, \bRtot, \etabar, \Delta)p(\hnoise|\lambda, \bRtot, \etabar, \Delta)\right].
	\label{eq:derive7}
\end{align}
The second term of \ref{eq:derive7}, $p(\hsignal| \lambda, \bRtot, \etabar, \Delta)$, corresponds to the expected fraction of astrophysical candidates in the total number of candidates:
\begin{equation}
	p(\hsignal| \lambda, \bRtot, \etabar, \Delta)=p(\hsignal| \etabar, \Delta)=\etabar.
	\label{eq:frac_signal}
\end{equation}
The last term in \eqref{eq:derive7}, $p(\hnoise|\lambda, \bRtot, \etabar, \Delta)$, is the expected fraction of noise-originating candidates, and it is mutually exclusive with $p(\hsignal| \etabar, \Delta)$, then
\begin{equation}
	p(\hnoise| \lambda, \bRtot, \etabar, \Delta)=p(\hnoise| \etabar, \Delta)=1 - \etabar.
\end{equation}
As we define $\lambda$ as the set of all parameters that determine the shape of \ac{CBC} population distribution above, the first term of \eqref{eq:derive7} can be simplified as, $p(x|\hsignal, \lambda, \bRtot, \etabar, \Delta) = p(x|\hsignal, \lambda, \Delta)$, and it is the single-candidate likelihood under the signal hypothesis, or the \ac{PDF} of ranking statistic in the signal population.
The third term in \eqref{eq:derive7} is the single-candidate likelihood under the background hypothesis, and is independent of \acp{CBC}' population parameters: $p(x|\hnoise, \lambda, \bRtot, \etabar, \Delta)=p(x|\hnoise, \Delta)$.
Then, we obtain the hierarchical likelihood of the following form:
\begin{multline}
	p(D| \lambda, \bRtot, \etabar, \Delta)
	\propto\bRtot^{\Ntot}\ee^{-\bRtot} \\ \prod^{\Ntot}_{i=1}
	\left[\etabar p(x_i|\hsignal, \lambda, \Delta)
	+(1-\etabar) p(x_i|\hnoise, \Delta)\right].
	\label{eq:finalform}
\end{multline}
Substituting \eqref{eq:simplified_prior} and \eqref{eq:finalform} into the marginalization expression in \eqref{eq:bayes_theorem} shows that the term depending on $\bRtot$ no longer depends on the population shape parameters $\lambda$:
\begin{widetext}
\begin{align}
	p(\lambda|D, \Delta)
	&\propto \left[\int_0^{\infty}\diff{\bRtot}\bRtot^{\Ntot}\ee^{-\bRtot}\pi(\bRtot)\right]
	\int_0^1\diff{\etabar}\left[\etabar p(x_i|\hsignal, \lambda, \Delta)+(1-\etabar) p(x_i|\hnoise, \Delta)\right]\pi(\lambda)\pi(\etabar),\\
	&\propto \int_0^1\diff{\etabar}
	\left[\etabar p(x_i|\hsignal, \lambda, \Delta)+(1-\etabar) p(x_i|\hnoise, \Delta)\right]\pi(\lambda)\pi(\etabar).
\end{align}
\end{widetext}
As a result, our parameter estimation can be restricted to the population shape parameters $\lambda$ and the signal fraction $\etabar$.
It is therefore sufficient to consider the hierarchical likelihood
\begin{multline}
	p(D| \lambda, \etabar, \Delta) \propto \\ \prod^{\Ntot}_{i=1}
	\left[\etabar p(x_i|\hsignal, \lambda, \Delta)+(1-\etabar) p(x_i|\hnoise, \Delta)\right],
	\label{eq:reduced_hl}
\end{multline}
instead of $p(D| \lambda, \bRtot, \etabar, \Delta)$ in \eqref{eq:finalform}.

The difference between $p(x|\hsignal, \lambda)$ and $p(x|\hsignal, \lambda, \Delta)$ is basically just the normalization threshold.
From the Bayes' theorem,
\begin{equation}
	p(x|\hsignal, \lambda, \Delta) = \frac{p(\Delta|x, \hsignal, \lambda)p(x|\hsignal, \lambda)}{p(\Delta|\hsignal, \lambda)},
\end{equation}
where
\begin{equation}
	p(\Delta|x, \hsignal, \lambda) = p(x\geq\rsth|x, \hsignal, \lambda)
		=\left\{
		\begin{matrix}
			0 &x<\rsth,\\
			1 &x\geq\rsth,
		\end{matrix}
		\right.
		\label{eq:tukey}
\end{equation}
and
\begin{equation}
	p(\Delta|\hsignal, \lambda) = p(x\geq\rsth|\hsignal, \lambda)
	= \int_\rsth^\infty\diff{x} p(x|\hsignal, \lambda).
	\label{eq:det_prob}
\end{equation}
Therefore, it is found that $p(x|\hsignal, \lambda, \Delta)$ is normalized over the range in which $x$ satisfies the detection criteria, while $p(x|\hsignal, \lambda)$ is normalized over the entire range of possible values of $x$ (when considering the log-\ac{LR} of GstLAL, $x=\ln\lr$, $-\infty\leq\ln\lr\leq\infty$).
The same goes for $p(x|\hnoise)$ and $p(x|\hnoise, \Delta)$.
The probability described in \eqref{eq:det_prob} represents the fraction of detectable signals, and corresponds to the selection function, $\alpha(\lambda)$, in the previous studies \cite{Mandel_2019, Vitale_2021, Heinzel_2023}.
It generally depends on the population parameters $\lambda$, and therefore accounts for selection effects when inferring them.

We have now identified three essential ingredients required to compute the hierarchical likelihood:
\begin{itemize}
	\item the single-candidate likelihood under the background hypothesis: $p(x|\hnoise, \Delta)$ and
	\item the single-candidate likelihood under the signal hypothesis: $p(x|\hsignal, \lambda, \Delta)$.
\end{itemize}
In the following two subsections describe in turn how each of these ingredients can be modeled.

\subsection{Single-candidate likelihood under the background hypothesis: \texorpdfstring{$p(x|\hnoise, \Delta)$}{p(x|hnoise, Delta)}}\label{subsec:px_noise}
The \ac{PDF} of the ranking statistic for noise-originating candidates, $p(x|\hnoise)$, is among the data products produced by a detection pipeline, and used for \ac{FAR} determination \cite{Cannon_2013, Messick_2017, gstlal}:
\begin{equation}
	\text{FAR}(x) = \frac{M\int_x^\infty\diff{x^\prime} p(x^\prime|\hnoise, \Delta)}{T},
	\label{eq:far}
\end{equation}
where the factor, $M$, in \eqref{eq:far} is the number of noise-like candidates with the detection statistic above the normalization threshold of $p(x|\hnoise, \Delta)$.
Considering the case of $x=\ln\lr$, $p(x|\hnoise)=p(\ln\lr|\hnoise)$ is obtained by integrating the GstLAL noise model for all $n$ parameters over $n-1$ dimensional surfaces of constant $\ln\lr$ \cite{Cannon_2013, cannon2015likelihoodratiorankingstatisticcompact}.
Since the forms of the surfaces of constant $\ln\lr$ is not known, the integrals are approximated using importance-weighted sampling (See \cite[Section IV]{cannon2015likelihoodratiorankingstatisticcompact} for detail).
The procedure can be summarized as follows:
\begin{algorithm}
\caption{Noise model marginalization}
	\label{alg:noise_marg}
	\For{$i \leftarrow 1$ \KwTo $N_\mathrm{sample}$}{
		Draw a sample from the proposal: $\Theta_i \sim p(\Theta)$\;
		Compute $x \gets \ln\lr(\Theta_i)$\;
		Update histogram: $h_\text{noise}(x) += p(\Theta_i\mid \hnoise)/p(\Theta_i)$\;
}
\end{algorithm}

Regarding the fidelity of the GstLAL noise model, previous studies have shown good agreement between the model and the observed candidates distribution in the low-\ac{LR} (or low-\ac{IFAR}) region \cite{GW150914, BBHs_in_O1, Ewing_2024}, although they diverge in high-\ac{LR} region because of the presence of astrophysical signals.

\subsection{Single-candidate likelihood under the signal hypothesis: \texorpdfstring{$p(x|\hsignal, \lambda)$}{p(x|hsignal, lambda)}}\label{subsec:rankcosmo}
$p(x|\hsignal, \lambda)$ is the likelihood of observing a ranking statistic value $x$ under the assumption that the candidate originates from a \ac{GW} signal produced by a \ac{CBC}, with the source population described by the parameter $\lambda$.
It can also be regarded as the \ac{PDF} of the ranking statistic in the signal population.
Unlike $p(x|\hnoise)$ above, it is parameterized by $\lambda$.
Thus, we need to model this \ac{PDF} using the data products of the detection pipeline.

As described in \secref{sec:gstlal_lr}, the GstLAL \ac{LR} ranking statistic is determined by the $n$-dimensional parameter set describing each candidate, and is designed as a measure of how likely a given candidate is to be a \ac{GW} signal from a \ac{CBC} \cite{cannon2015likelihoodratiorankingstatisticcompact, Messick_2017, Tsukada_2023, gstlal}.
Since the $n$-dimensional set includes the \ac{SNR} and the best match template of the candidate of interest, in order to predict the distribution of the ranking statistic in the signal population, we need to assume the astrophysical and cosmological population models, which describe the distribution of the intrinsic parameters (masses, spins, etc.) and the luminosity distances of compact binaries.
In the GstLAL signal model, the \ac{CBC} rate is assumed to scale with the cube of the luminosity distance (in the local universe, this is equivalent to a uniform distribution in comoving volume) \cite{cannon2015likelihoodratiorankingstatisticcompact}, and the model of the \ac{CBC} intrinsic parameters distribution (\textit{mass model}) is given by the analyst.
In principle, the true population model is unknown---it is what we aim to learn from \ac{GW} observations.
Thus, any population model adopted within GstLAL cannot be correct in a strict sense.
Nevertheless, assuming some form of population model is an essential step in the detection process.

The parameterization of the population model employed in the GstLAL signal model may differ from the parameterization we wish to use in population analysis; even if the same parameterization is adopted, the population parameters are fixed to specific values.
Therefore, in order to perform population inference, we must reparameterize the GstLAL signal model in a way that aligns with our objectives.
It can be done as follows:
\begin{align}
	&p(x|\hsignal,\lambda)
	=\int\diff{\rcomb}\diff{\template}p(x,\rcomb, \template|\hsignal, \lambda)\\
	&=\int\diff{\rcomb}\diff{\template}p(x| \rcomb, \template, \hsignal, \lambda)
	p(\rcomb, \template| \hsignal, \lambda)\\
	&=\int\diff{\rcomb}\diff{\template}p(x| \rcomb, \template, \hsignal)
	p(\rcomb, \template| \hsignal, \lambda).
	\label{eq:signalmodel_reparametorization}
\end{align}
where $\rcomb$ denotes the combined \ac{SNR} of a network, and is defined by
\begin{equation}
	\rcomb=\sqrt{\sum_{\rho\in\vrho}\rho^2}.
	\label{eq:rcomb}
\end{equation}
The second term in \eqref{eq:signalmodel_reparametorization}, $p(\rcomb, \template| \hsignal, \lambda)$, determined by the sensitivity of the instruments of interest and the assumed population model (The discussion in \cite[Chapter 3]{Fong_thesis} might be useful).
The first term in \eqref{eq:signalmodel_reparametorization} depends only on the characteristic of the search pipeline.
It can be written as:
\begin{equation}
	p(x| \rcomb, \template, \hsignal) = \frac{p(x, \rcomb, \template| \hsignal)}{\int\diff{x}p(x, \rcomb, \template| \hsignal)},
\end{equation}
and the joint distribution of the log-\ac{LR} ranking statistic, the combined \ac{SNR} and the best-matched template, $p(x, \rcomb, \template| \hsignal)$, is obtained by marginalizing GstLAL signal model.
The marginalization can be done in the same sampling loop shown in \algref{alg:noise_marg}:
\begin{algorithm}
\caption{Noise and signal model marginalization}
	\label{alg:noise_signal_marg}
	\For{$i \leftarrow 1$ \KwTo $N_\mathrm{sample}$}{
		Draw a sample from the proposal: $\Theta_i \sim p(\Theta)$\;
		Compute $x \gets \ln\lr(\Theta_i)$\;
		Update histogram: $h_\text{noise}(x) += p(\Theta_i\mid \hnoise)/p(\Theta_i)$\;
		Compute $\rcomb \gets \rcomb(\vec{\rho_i})$\;
		Update histogram: $h_\text{signal}(x, \rcomb, \template_i)+=p(\Theta_i\mid \hsignal)/p(\Theta_i)$\;
}
\end{algorithm}

Through the method described above, we can in principle replace the mass model and the spatial distribution model employed in the GstLAL signal model with the population model which we wish to use.
However, in the importance-weighted sampling procedure, increasing the dimensionality of the target distribution requires a larger number of effective samples, which in turn leads to a higher computational cost.
Since the primary aim of this work is not to obtain a scientifically meaningful estimate of cosmological parameters from real data, but rather to demonstrate the effectiveness of the proposed method, we perform the mock data analysis in an idealized setting in \secref{sec:mda}.
Specifically, we assume that the true intrinsic parameters distribution of compact binaries is already perfectly known and that the mass model in the GstLAL signal model is set to be consistent with the true distribution.
Under this assumption, there is no need to replace the mass model in the detection pipeline.
The reparametorization in \eqref{eq:signalmodel_reparametorization} then simplifies:
\begin{align}
	&p(x|\hsignal,\lambda)
	=\int\diff{\rcomb}p(x,\rcomb|\hsignal, \lambda)\\
	&=\int\diff{\rcomb}p(x| \rcomb, \hsignal, \lambda)
	p(\rcomb| \hsignal, \lambda)\\
	&=\int\diff{\rcomb}p(x| \rcomb, \hsignal)
	p(\rcomb| \hsignal, \lambda).
	\label{eq:signalmodel_reparametorization2}
\end{align}
The first term in \eqref{eq:signalmodel_reparametorization2}, $p(x| \rcomb, \hsignal)$, denotes the likelihood that a ranking statistic value $x$ is assigned to a candidate with combined (network) \ac{SNR}, $\rcomb$.
It is a conditional probability distribution incorporating the characteristics of the detection pipeline, which can be obtained using the importance-weighted sampling procedure outlined above.
The second term in \eqref{eq:signalmodel_reparametorization2} determined by the sensitivity of the instruments of interest, the mass distribution of \ac{CBC} in the universe, and our cosmological model.

\subsubsection{Modeling PDF of single detector SNR}\label{subsubsec:single_det_snr_dist}
In this subsection, we consider the calculation of the last term of \eqref{eq:signalmodel_reparametorization2} for a single detector case, $\rcomb = \sqrt{\rho^2} = \rho$.
We explicitly denote $\rho$ as $\robs$ in this subsection to clarify that it refers to the observed \ac{SNR}.

We define the optimal \ac{SNR}, $\ropt$, as the expected \ac{SNR} of a \ac{GW} from a \ac{CBC} which is optimally oriented and located to maximize the strain amplitude at a given detector.
When we denote the redshift and the set of intrinsic source parameters that determine the value of $\ropt$ by $z$ and $\theta$, respectively, the last term of \eqref{eq:signalmodel_reparametorization2} can be written as:
\begin{align}
	&p(\robs|\hsignal, \lambda) = \int\diff{\theta}\diff{z}p(\robs, \theta, z|\hsignal, \lambda)\\
	&=\int\diff{\theta}\diff{z}p(\robs|\theta, z, \hsignal, \lambda)p(\theta, z|\hsignal, \lambda)\\
	&=\int\diff{\theta}\diff{z}p(\robs|\ropt(\theta, z, \lc), \hsignal)\nonumber\\
	&\hspace{35mm}\times p(\theta|\hsignal, \lm)p(z|\hsignal, \lc).
	\label{eq:introducevTheta}
\end{align}
The first term of \eqref{eq:introducevTheta} depends on the cosmological parameters, $\lc$, and on our cosmological model thorough the conversion from redshift to luminosity distance.
In addition, since $\robs$ here is the observed \ac{SNR}, it should incorporate the measurement uncertainty of \ac{SNR} caused by detector noise realization and the randomness of the sky location and orbital inclination of the source.
The second term of \eqref{eq:introducevTheta} is determined by the astrophysical population.
The third term can be constructed either on the basis of our cosmological model or, as commonly done in previous studies, using existing galaxy catalogues \cite{GW170817_wo_em, Soares_santos_2019, Gray_2020, gwtc-3_cosmo, gwtc-4_cosmo, Petiteau_2011, icarogw, Beirnaert_2025}.

\subsubsection{Measurement uncertainty of SNR}
As mentioned above, the \ac{SNR} of a \ac{GW} signal from a compact binary source, characterized by a specific intrinsic parameter and a redshift, should scatter randomly around the optimal \ac{SNR}, $\ropt$, defined earlier.
This randomness arises from the realization of instrument noise as well as from the random sky location and inclination angle of the source, and such uncertainties must be incorporated into the first term of \eqref{eq:introducevTheta}.
In what follows, we examine how this uncertainty can be incorporated into the first term of \eqref{eq:introducevTheta}, and how that term can be expressed more explicitly in terms of the optimal \ac{SNR}, $\ropt$.
\begin{figure*}
	\resizebox{0.9\linewidth}{!}{\includegraphics{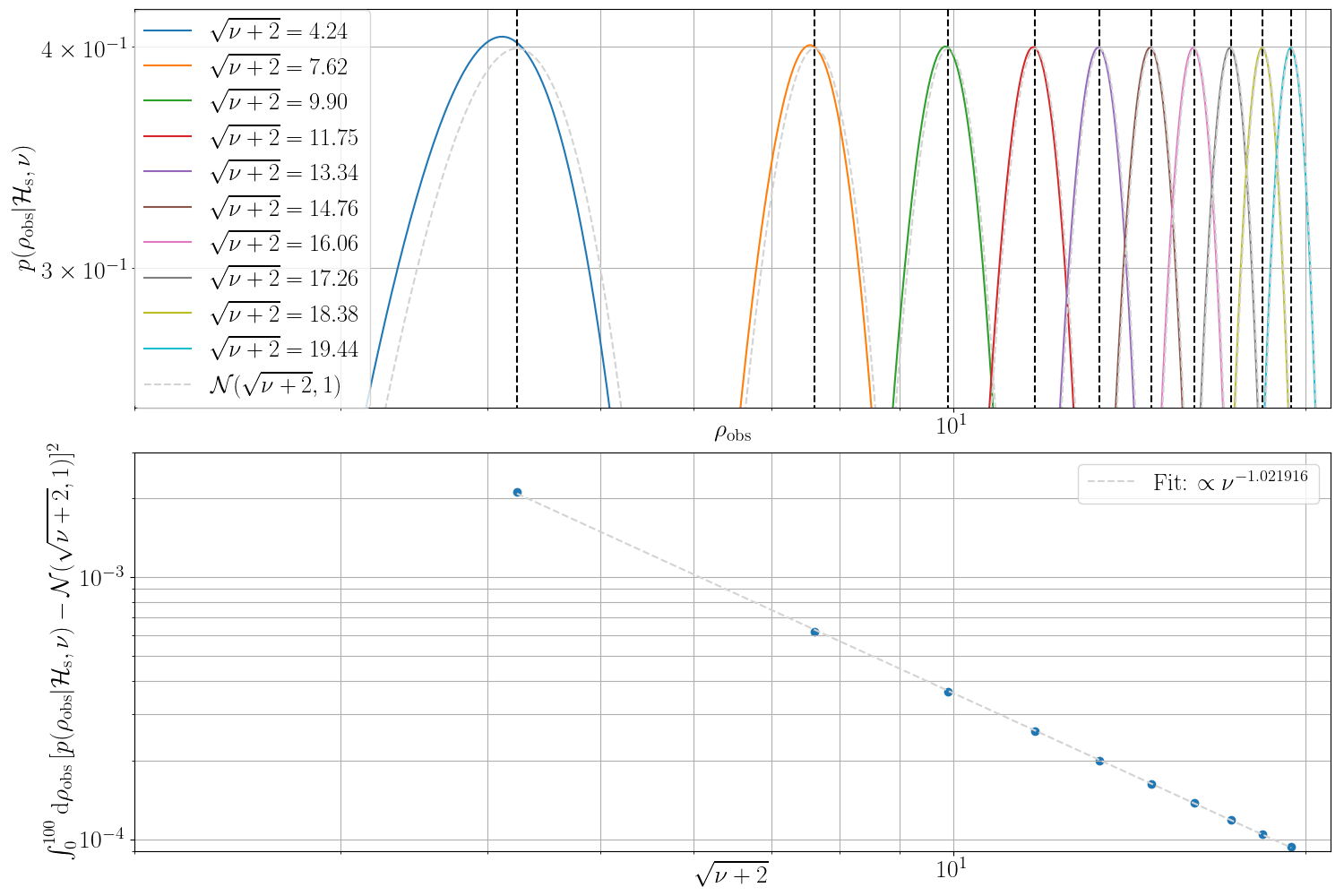}}
	\caption{Comparison between the unit-variance Gaussian distribution, $\gaussian(\robs|\sqrt{\nu+2}, 1)$, and the expected distribution of the observed \ac{SNR} $\robs$, $p(\robs|\hsignal, \nu)$, derived from the non-central chi-squared distribution, conditioned by the non-centrality $\nu = (\ropt\mG)^2 - 2$.
	\textit{Top panel}: the two distributions are shown for different value of $\nu$; the colored solid lines represent the exact $p(\robs|\hsignal, \nu)$, while the gray dashed lines represent the unit-variance Gaussian distributions.
	The two distributions converge in the high-\ac{SNR} limit.
	\textit{Bottom panel}: the mean-squared error between $p(\robs|\hsignal, \nu)$ and the unit-variance Gaussian distribution as a function of $\sqrt{\nu + 2} = \ropt\mG$, showing an $\sim\nu^{-1}$ scaling.}
	\label{fig:snr_error}
\end{figure*}

First, we consider the randomness caused by instrumental noise.
As described in the section 7.8.1 of \cite{Creighton_2011}, matched filtering algorithms for \ac{CBC} have traditionally filtered the data against a set of complex template, and the observed \ac{SNR}, $\robs$, is defined as the modulus of the complex \ac{SNR}.
When we assume the zero-mean stationary Gaussian noise and normalized templates, the real and imaginary part of the complex \ac{SNR} become unit-variance Gaussian random variables \cite[Section 7.2.4]{Creighton_2011}.
Therefore, under these assumptions, the distribution of the squared observed \ac{SNR}, $\Xobs\equiv\robs^2$, becomes the non-central $\chi^{2}$ distribution with two degrees of freedom \cite{non-central-chi-square}:
\begin{equation}
	p(\Xobs|\hsignal, \nu) = \frac{1}{2}\ee^{-\frac{\Xobs+\nu}{2}}I_{0}(\sqrt{\nu \Xobs}),
	\label{eq:ncx2}
\end{equation}
where $\nu$ is the non-centrality parameter which is related to the expected value of the $\Xobs$, and $I_0$ is the modified Bessel function of the first kind of order zero.
When $\Xexp$ denotes the expected value of $\Xobs$,
\begin{equation}
	\Xexp = 2 + \nu.
\end{equation}
It is reasonable to assume that
\begin{equation}
	\Xexp = \left(\ropt\mG\right)^2,
\end{equation}
where $\mG$ is defined as the geometry factor of the instrument expressed as:
\begin{equation}
	\mG = \sqrt{{F_+(\vec{n}, \psi)}^2 \left(\frac{1+\cos^2\iota}{2}\right) + {F_\times(\vec{n}, \psi)}^2\cos^2 \iota}.
\end{equation}
Here, $\iota$ is the orbital inclination, and $F_+$ and $F_\times$ are the antenna pattern functions \cite{Creighton_2011}.
They are determined by the unit vector, $\vec{n}$, pointing to the location of the \ac{CBC} source on the sky and the polarization angle, $\psi$.
From
\begin{equation}
	p(\robs) \diff\robs = p(\Xobs)\diff\Xobs = 2\robs p(\Xobs)\diff\robs,
\end{equation}
we find that the first term of \eqref{eq:introducevTheta} can be written as:
\begin{align}
	&p(\robs|\ropt, \hsignal) = \int_0^1 \diff{\mG} p(\robs, \mG|\ropt, \hsignal)\\
	&= \int_0^1 \diff{\mG} p(\robs|\mG, \ropt, \hsignal)p(\mG|\hsignal)
	\label{eq:probsconditionedropt0}
	\\
	&= \int_0^1 \diff{\mG} p(\mG|\hsignal)\nonumber\\
	&\hspace{5mm}\times 2\robs\, p\left(\Xobs = \robs^2 \middle|\hsignal, \nu = (\ropt\mG)^2 -2\right).
	\label{eq:probsconditionedropt1}
\end{align}
The last term of \eqref{eq:probsconditionedropt1} can be calculated using \eqref{eq:ncx2}, and the first term, $p(\mG|\hsignal)$, can be obtained from simulation assuming a uniform and isotropic distribution of gravitational wave sources.

As shown in \figref{fig:snr_error}, in the high-\ac{SNR} limit the distribution, $p(\robs|\mG, \ropt, \hsignal)$, in \eqref{eq:probsconditionedropt0} and \eqref{eq:probsconditionedropt1} approaches a Gaussian distribution, $p(\robs| \mu, \sigma)=\frac{1}{\sqrt{2\pi\sigma^2}}\exp[-(x-\mu)^2 / 2\sigma^2]$, with the mean of $\mu=\ropt\mG$ and the unit variance, $\sigma=1$.
Even at $\ropt\mG\simeq 4$, the mean-squared error is only about 0.1\%.
Motivated by this, in the \acp{MDA} presented in \secref{sec:mda}, we simplify the treatment by replacing $p(\robs|\mG, \ropt, \hsignal)$ with the unit-variance Gaussian distribution:
\begin{equation}
	p(\robs|\ropt, \hsignal)
	\simeq \int_0^1 \diff{\mG} p(\mG|\hsignal)\gaussian(\robs|\ropt\mG, 1).
	\label{eq:probsconditionedropt2}
\end{equation}

\subsubsection{PDF of combined SNR}
\label{subsubsec:combined_snr_pdf}

In this subsection, for simplicity, we assume a detector network composed of multiple instruments with nearly identical sensitivities---such as the two LIGO detectors.
Under this assumption, we show how to construct the distribution of the combined \ac{SNR} from the single detector \ac{SNR} distribution and the distribution of the relative magnitude of geometry factors of different instruments.
Although in reality some candidates are observed by only a subset of the $\Ndet$ instruments, for simplicity we assume that all candidates in our dataset $D$ are identified by all $\Ndet$ detectors and examine the corresponding distribution of the combined \ac{SNR}.

Then, for now, it is assumed that we have $\Ndet = m + 1$ instruments ($m$ is an integer equal to or greater than 1), and we already know the distribution of \acp{SNR} observed by the 0th instrument, $O_0$, for the given value of cosmological parameters, $p(\rho_0|\hsignal, \lambda)$.
When the ratio of \ac{SNR} observed by 0th and $i$th instruments for a given source is written as
\begin{equation}
	r_{i, 0} \equiv \frac{\rho_i}{\rho_0},
\end{equation}
the combined \ac{SNR} defined by \eqref{eq:rcomb} can be expressed as
\begin{equation}
	\rcomb = \rho_{0} \sqrt{1 + \sum_{i=1}^{m}r_{i, 0}^2}.
\end{equation}
Considering
\begin{multline}
	p(\rho_0, r_{1, 0}, \cdots, r_{m, 0})\diff{\rho_0}\prod_{i=1}^{m}\difff{r_{i,0}} =\\
	p(\rcomb, r_{1, 0}, \cdots, r_{m, 0})\diff{\rcomb}\prod_{i=1}^{m}\difff{r_{i,0}}
\end{multline}
and
\begin{equation}
	\difff{\rcomb} = \left(\sqrt{1 + \sum_{i=1}^{m}r_{i, 0}^2}\right)\difff{\rho_0},
\end{equation}
we obtain
\begin{align}
	&p(\rcomb|\hsignal, \lambda)\nonumber\\
	&=\int \left(\prod_{i=1}^{m}\difff{r_{i,0}}\right)\,p(\rcomb, r_{1, 0}, \cdots, r_{m, 0}|\hsignal, \lambda)\\
	&=\int \left(\prod_{i=1}^{m}\difff{r_{i,0}}\right)\,\frac{1}{\sqrt{1 + \sum_{i=1}^{m}r_{i, 0}^2}}\nonumber\\
	&\hspace{3mm}\times p\left(\rho_0 = \frac{\rcomb}{\sqrt{1 + \sum_{i=1}^{m}r_{i, 0}^2}}, r_{1, 0}, \cdots, r_{m, 0}\middle|\hsignal, \lambda\right).
	\label{eq:rcombdist}
\end{align}

Since the distribution of $r_{i, 0}$ depends roughly only on the geometric properties of the instrument network and does not depend on cosmological parameters or astrophysical mass distribution models, the last term in \eqref{eq:rcombdist} can be factorized as follows:
\begin{multline}
	p(\rho_0, r_{1, 0}, \cdots, r_{m, 0}|\hsignal, \lambda) = \\
	p(\rho_0|\hsignal, \lambda)p(r_{1, 0}, \cdots, r_{m, 0}|\hsignal).
\end{multline}
When the sensitivity of two detectors are very similar, the ratio of the \acp{SNR} between them is nearly equal to the ratio of their geometry factors,
\begin{equation}
	r_{i, 0} \simeq \frac{\mG_i}{\mG_0}.
\end{equation}
Accordingly, the approximated distribution of $r_{i, 0}$ can be obtained by computing the geometry factors for each instrument over a set of isotropically distributed sky locations and uniformly distributed inclinations, and taking their ratios.

\subsection{Point estimate of signal fraction}\label{subsec:fraction_estimate}
This subsection details the procedure for obtaining a point estimate of the signal fraction, $\etabar$, which was mentioned around \eqref{eq:dirac_delta_prior}.
Recalling \eqref{eq:frac_signal}, the signal fraction can be written using the probability of astrophysical origin, $p(\hsignal|x, \lambda, \etabar, \Delta)$, so-called $\pastro$:
\begin{equation}
	\etabar = p(\hsignal|\lambda, \etabar, \Delta)= \int\diff{x} p(\hsignal|x, \lambda, \Delta)p(x|\lambda, \etabar, \Delta).
	\label{eq:frac_signal_pastro}
\end{equation}
Here we define $p(x|\text{observed})$ as the observed distribution of the detection statistic, and using the true value of the signal fraction $\etabar_0$ it can be written as
\begin{equation}
	p(x|\text{observed}) = \etabar_0 p(x|\hsignal, \lambda, \Delta) + (1 - \etabar_0) p(x|\hnoise, \Delta),
	\label{eq:def_p_observe}
\end{equation}
if our population model is correct and the total number of observed candidates $\Ntot$ is sufficiently large.
Assuming
\begin{equation}
	p(x|\lambda, \etabar, \Delta)\simeq p(x|\text{observed}),\label{eq:likelihood_approx}
\end{equation}
we obtain an approximation of the signal fraction, and we refer to it as $\etabar_1^\prime$:
\begin{align}
	\etabar&\simeq \int\diff{x} p(\hsignal|x, \lambda, \etabar, \Delta)p(x|\text{observed})\label{eq:approx_integral}\\
	&\simeq \frac{1}{\Ntot}\sum_{i=1}^{\Ntot}p(\hsignal|x_i, \lambda, \Delta)\equiv \etabar_1^\prime,
	\label{eq:pastro_sum}
\end{align}
Since the calculation of $p(\hsignal|x, \lambda, \etabar, \Delta)$ requires us to know $\etabar$ as shown later, we approximate $p(\hsignal|x, \lambda, \etabar, \Delta)$ in \eqref{eq:approx_integral} with the $\pastro$ marginalized over $\etabar$:
\begin{align}
	p(\hsignal|x, \lambda, \Delta) &= \int_0^1\diff{\etabar}p(\hsignal, \etabar|x, \lambda, \Delta)\\
	&= \int_0^1\diff{\etabar}p(\hsignal|x, \lambda, \etabar, \Delta)p(\etabar|\lambda, \Delta).
	\label{eq:pastro}
\end{align}
From Bayes' theorem,
\begin{align}
	&p(\hsignal|x, \lambda, \etabar, \Delta) = \nonumber\\
	&\frac{p(\hsignal| \lambda, \etabar, \Delta)p(x|\hsignal, \lambda, \Delta)}{p(\hsignal| \lambda, \etabar, \Delta)p(x|\hsignal, \lambda, \Delta)+ p(\hnoise| \lambda, \etabar, \Delta)p(x|\hnoise, \Delta)}\\
	&= \frac{\etabar p(x|\hsignal, \lambda, \Delta)}{\etabar p(x|\hsignal, \lambda, \Delta)+ (1 - \etabar)p(x|\hnoise, \Delta)}.
	\label{eq:pastro_not_marginalized}
\end{align}
$p(x|\hnoise, \Delta)$ and $p(x|\hsignal, \lambda, \Delta)$ is the \ac{PDF} of the detection statistic in the noise and signal population, and these are described in the previous subsections.
We adopt the uniform prior between 0 and 1 as $p(\etabar|\lambda, \Delta)$.

If $\Ntot$ is sufficiently large and the approximation in \eqref{eq:likelihood_approx} holds well, the fraction of signals is expected to be estimated with adequate accuracy using \eqref{eq:pastro_sum} (as well as \eqref{eq:pastro} and \eqref{eq:pastro_not_marginalized}).
In the \acp{MDA} in \secref{sec:mda}, however, we find that this approximation does not works well especially in the case of the smaller injected values of $\etabar$, and moreover that the resulting discrepancy can have a serious impact on the posterior distribution of the population parameters, which is the final outcome of our analysis.
On the other hand, we also found that the true signal fraction, $\etabar_0$, and $\etabar_1^\prime$ in \eqref{eq:pastro_sum} are related by a linear relation:
\begin{equation}
	\etabar_1^\prime \simeq (F - B) \etabar_0 + B,
	\label{eq:true_estimate_relation}
\end{equation}
where
\begin{equation}
	F(\lambda, \Delta) \equiv \int_{-\infty}^\infty\diff{x} p(\hsignal|x, \lambda, \Delta)p(x|\hsignal, \lambda, \Delta)
	\label{eq:pastro_exp_signal}
\end{equation}
and
\begin{equation}
	B(\lambda, \Delta) \equiv \int_{-\infty}^\infty \diff{x}p(\hsignal|x, \lambda, \Delta)p(x|\hnoise, \Delta).
	\label{eq:pastro_exp_noise}
\end{equation}
The linear relation in \eqref{eq:true_estimate_relation} can be readily derived by substituting \eqref{eq:def_p_observe} into \eqref{eq:approx_integral}.
We define the point estimate corrected using the relation in \eqref{eq:true_estimate_relation} as
\begin{equation}
	\etabar_0^\prime \equiv \frac{\etabar_1^\prime - B}{F - B}.
	\label{eq:def_gammabar_0_prime}
\end{equation}
The discovery of the relation in \eqref{eq:true_estimate_relation} enables us to obtain more accurate estimates of the signal fraction from the observational dataset, which can then be incorporated into the inference of the population parameter.
The comparison between the results of inferences employing different methods of estimation of signal fraction is shown in \secref{subsubsec:results_using_point_estimate}.

	As may already be apparent from the preceding discussion, the order of presentation above reflects the heuristic path by which we arrived at the estimator in \eqref{eq:def_gammabar_0_prime}, rather than the minimal conceptual structure required to understand it.
	In fact, for the conceptual understanding of the estimator, the probabilistic derivation in \eqref{eq:frac_signal_pastro} and \eqref{eq:approx_integral} is not essential.
	What truly underlies the estimator is the fact that the expectation value of any function of the detection statistic differs between the signal and noise populations, and that the observed mean reflects a mixture of the two according to the true signal fraction.
	The $\pastro$-based formulation simply provides a convenient and interpretable way to express this relation, and this choice was also empirically justified in our \ac{MDA}, as discussed in \secref{subsec:mda_results}.

\section{Proof-of-concept Mock data analyses}\label{sec:mda}
In this section, we validate our proposed method by performing simplified, proof-of-concept \acp{MDA}.
In our proof-of-concept \acp{MDA}, we did not inject signals into detector strain and run the detection pipeline to generate a list of \ac{GW} candidates.
Instead, we drew fixed numbers of ranking statistic values corresponding to candidates from our noise and signal models, respectively, and used this synthetic (fake) dataset to compute the posterior distribution of $\ho$.
Accordingly, in a \ac{MDA} our dataset consists of a list of $\Ntot$ values of the log-likelihood ratio.
Assuming that all population parameters other than the Hubble constant, $\ho$, are known, we investigate whether the injection value of $\ho$ can be recovered from a dataset consisting of $\Ntot$ fake candidates.
We use the data products generated in the S5S6 reanalysis \cite{s5s6_2021_reanalysis} for the fake data generation and the construction of the single-candidate likelihood under the signal/noise hypothesis.

The population model employed in our \acp{MDA} and the single-candidate likelihood under the signal hypothesis based on it are described in \secref{subsec:mda_pop_model}.
\secref{subsec:mda_data_generation} describes our fake data generation process.
In \secref{subsec:mda_results}, we show the example posterior and the results of the \ac{p-p} test.

\begin{figure*}
	\resizebox{\linewidth}{!}{\includegraphics{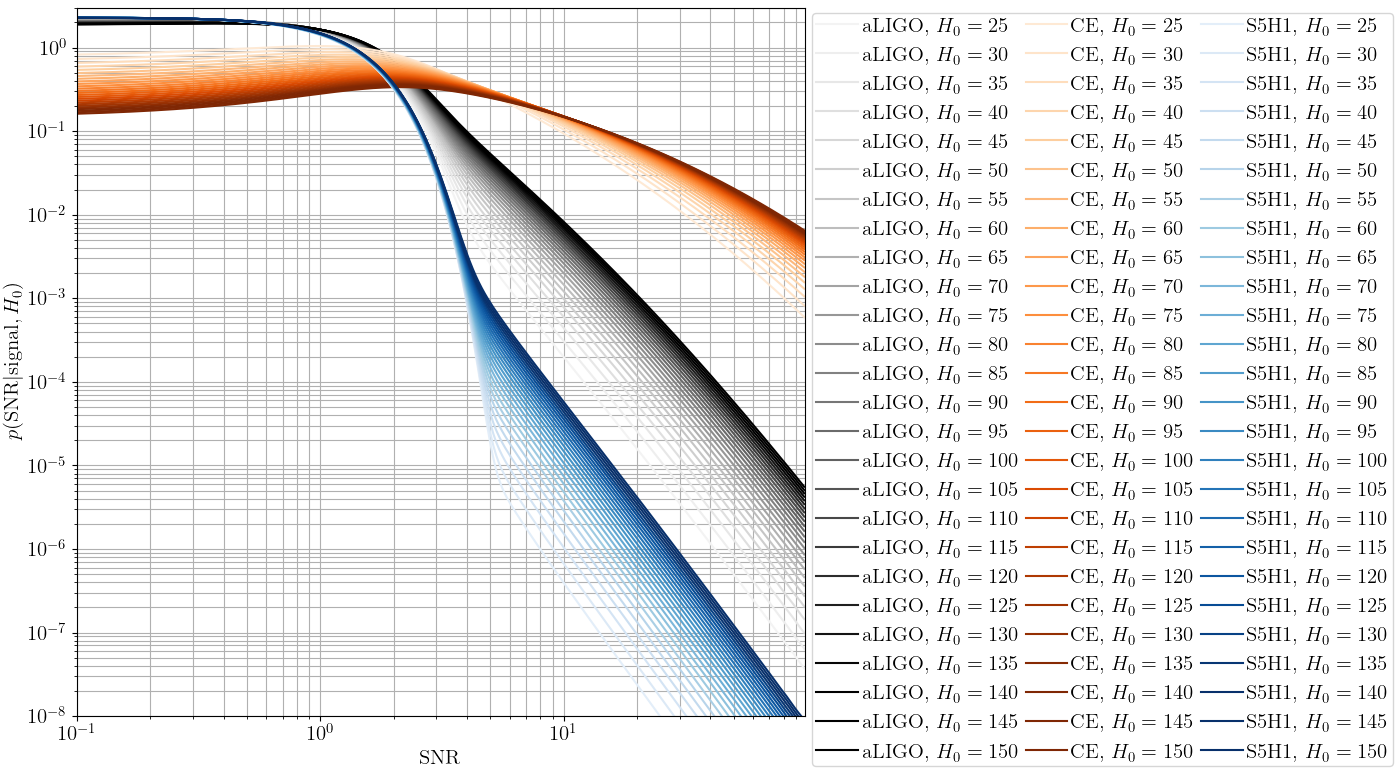}}
	\caption{Expected \ac{SNR} distribution assuming a given instrument and the population model described in \secref{subsec:mda_pop_model}.
	The gray lines, orange lines and blue lines correspond to Cosmic Explorer \cite{LIGO-P1600143}, Advanced LIGO design sensitivity \cite{LIGO-T1800044}, and S5 sensitivity, spanning GPS time \mdagpstime, respectively.
	}
	\label{fig:expected_snr_dist}
\end{figure*}
\begin{figure*}
\subfloat[\label{subfig:signalmodel_given_snr}$\ln \lr$ distribution model under the signal hypothesis conditioned by observed \ac{SNR}, obtained by marginalizing GstLAL signal model assuming S5 sensitivity via importance sampling described in \algref{alg:noise_signal_marg}.]{\resizebox{.5\linewidth}{!}{\includegraphics{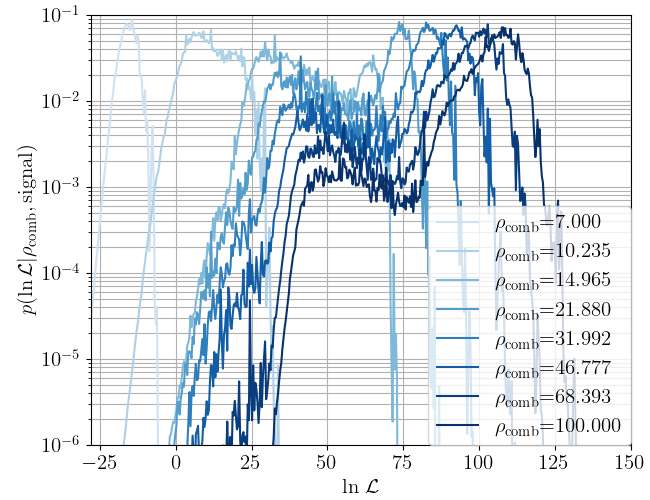}}}%
\subfloat[\label{subfig:signalmodel_given_H0}$\ln \lr$ distribution model under the signal hypothesis for a given value of $\ho$, obtained by integrating the results shown by blue lines in \figref{fig:expected_snr_dist} and \figref{subfig:signalmodel_given_snr}.  As a reference, the original marginalized GstLAL signal model is plotted in orange.]{\resizebox{.5\linewidth}{!}{\includegraphics{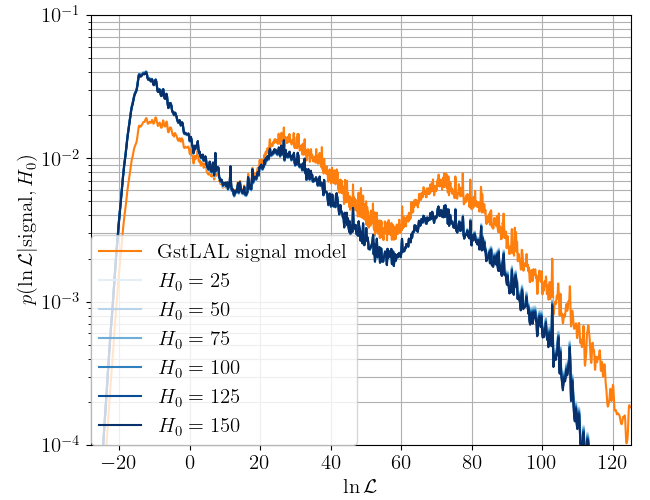}}}%
\caption{Detection statistic distribution models employed in the \ac{MDA}.  These \acp{PDF} were generated based on data products provided by a previous study \cite{s5s6_2021_reanalysis}.  The $\ho$-dependence of this \ac{MDA} signal model remains limited, since the LIGO detectors during the S5 run were not sensitive to very distant regions of the universe.}
\end{figure*}
\begin{figure*}
\subfloat[\label{subfig:signal_snr_hist}The combined \ac{SNR} distribution of the signal-originating candidates in an example \ac{MDA} dataset.  The black solid line shows the network \ac{SNR} distribution model corresponding to the injected value of $\ho$, $\ho=$\SI{70}{\kilo\metre\per\second\per\mega\parsec}, as a reference.]{\resizebox{.5\linewidth}{!}{\includegraphics{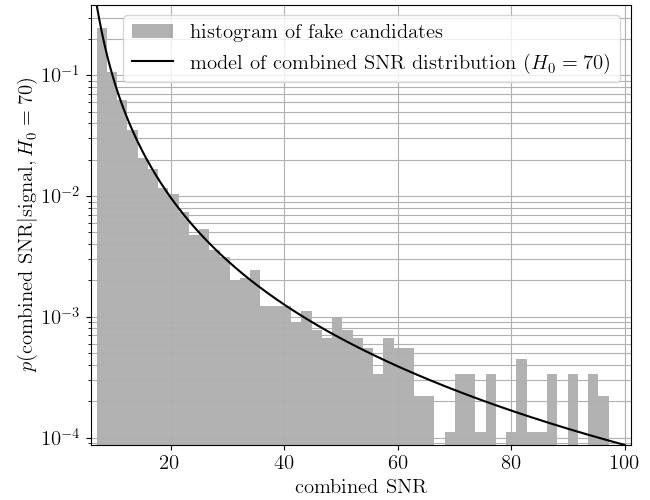}}}%
\subfloat[\label{subfig:lr_hist}The log-\ac{LR} distribution of the candidates in an example \ac{MDA} dataset.  The black dashed line presents the GstLAL noise model, and the black solid line shows the \ac{MDA} signal model corresponding to the injected value of $\ho$, $\ho=$\SI{70}{\kilo\metre\per\second\per\mega\parsec}.]{\resizebox{.5\linewidth}{!}{\includegraphics{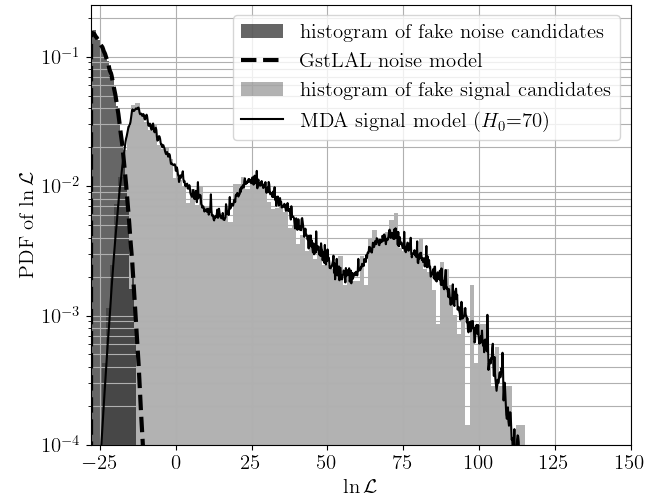}}}%
\caption{Comparison between the distribution models, and the distribution of candidates' \ac{SNR} and log-\ac{LR} in an example \ac{MDA} dataset.}
\label{fig:comp_sample_originalpdf}
\end{figure*}

\subsection{Population model}\label{subsec:mda_pop_model}
To perform the \ac{p-p} test, we prepared 1,386 mock universes, corresponding to 126 different values of the Hubble constant $\ho$ and 11 different values of the signal fraction $\etabar$.
For consistency with the prior described in \secref{subsec:framework_basics}, $\ho$ is varied uniformly from \SI{25}{\kilo\metre\per\second\per\mega\parsec} to \SI{150}{\kilo\metre\per\second\per\mega\parsec} in increments of \SI{1}{\kilo\metre\per\second\per\mega\parsec}, $\etabar$ was sampled uniformly from 0 to 1 with a step size of 0.1.

We begin by describing the \ac{CBC} intrinsic parameter distribution in the mock universes, which corresponds to the second term of \eqref{eq:introducevTheta}.
For simplicity, we consider only the lowest post-Newtonian order.
At this level, the optimal \ac{SNR} defined in \secref{subsubsec:single_det_snr_dist} depends only on the detector frame chirp mass, $\mchirpd$, and the luminosity distance between the source and the Earth, $\dl$: $\ropt = \ropt(\mchirpd, \dl)$.
The source frame and the detector frame chirp mass is defined as:
\begin{equation}
	\mchirps = \left(\frac{\pms^3\sms^3}{\pms + \sms}\right)^{\frac{1}{5}},
	\label{eq:source_frame_chirp}
\end{equation}
and
\begin{equation}
	\mchirpd = (1+z) \mchirps
	= \left(\frac{\pmd^3\smd^3}{\pmd + \smd}\right)^{\frac{1}{5}},
	\label{eq:conversion_Mcs_Mcd}
\end{equation}
where $\pms$ and $\sms$ are the source-frame masses of the binary components ($\pms\geq\sms$), and $\pmd=(1+z)\pms$ and $\smd=(1+z)\sms$ are the detector-frame masses of the binary components.
Defining $\mchirpd_{,8}$ and $\dl_{,8}$ such that $\ropt(\mchirpd_{,8}, \dl_{,8})= 8$, we can calculate $\ropt$ for a given set of values of chirp mass and luminosity distance as:
\begin{equation}
	\ropt(\mchirpd, \dl) = 8\left(\frac{\mchirpd}{\mchirpd_{,8}}\right)^{5/6}\frac{\dl_{,8}}{\dl}.\label{eq:ropt}
\end{equation}
$\mchirpd_{,8}$ and $\dl_{,8}$ depend on the sensitivity of the instrument of interest.
Our \acp{MDA} used the power spectral density estimated from a segment of LIGO Hanford data in S5, spanning GPS times \mdagpstime.
The reference distance $\dl_{,8}$ for a given $\mchirpd_{,8}$ was computed using the \texttt{gstlal.psd.HorizonDistance} function in the GstLAL package \cite{gstlal}, which evaluates the horizon distance for a binary with specified component masses and detector power spectral density.
Accordingly, in this setting, the only intrinsic parameter of the binary that controls the optimal \ac{SNR} is the chirp mass in the source frame, $\mchirps$.
This implies that the second term in \eqref{eq:introducevTheta} can be expressed as:
\begin{equation}
	p(\theta|\hsignal, \lm) = p(\mchirps|\hsignal, \lm).
	\label{eq:mda_mass_model}
\end{equation}
	In typical population studies of \ac{GW} sources, it is common to assume a source-frame mass distribution motivated by astrophysical considerations as in \eqref{eq:introducevTheta}.
In our \acp{MDA}, however, we choose to fix the mass distribution in the detector frame to maintain consistency with the run configuration used in \cite{s5s6_2021_reanalysis}, whose data products are used in this study.
Since the focus of our \acp{MDA} is not on extracting redshift information from features in the mass distribution, this choice is not important.
Considering $p(\mchirpd|z, \hsignal, \lm) = \frac{1}{(1 + z)}p(\mchirps|\hsignal, \lm)$, \eqref{eq:introducevTheta} can be rewritten as:
\begin{multline}
	p(\robs|\hsignal, \lambda) =\\
	\int\diff{\mchirpd}\diff{z}p(\robs|\ropt(\mchirpd, \dl(z, \lc)), \hsignal)\\
	\times p(\mchirpd|z, \hsignal, \lm)p(z|\hsignal, \lc).
	\label{eq:introducevThetarewrite}
\end{multline}
The binary detector-frame component masses are assumed to be log-uniformly distributed in the mock universes.
The mass range is set to be 1--$100\,\Msun$ for the detector-frame component masses and up to $100\,\Msun$ for the detector-frame total mass.
Then, we obtained $p(\mchirpd|z, \hsignal, \lm)$ numerically by drawing samples of $\pmd$ and $\smd$ from $p(m)\propto 1/m$, and calculating \eqref{eq:conversion_Mcs_Mcd} for each sample.

We now turn to the third term of \eqref{eq:introducevTheta}.
In the mock universes, compact binaries are uniformly distributed in the comoving volume, $\Vc$.
Therefore,
\begin{equation}
	p(z|\hsignal, \lc)\propto\frac{T}{1+z}\frac{\difff{\Vc}}{\difff{z}},
\end{equation}
where $T$ is the length of the observation in the detector frame ($T/(1+z)$ is the length of the observation in the source frame).
Furthermore, the mock universes are ``flat''.
Then the expressions of the comoving distance, $\dc$, and the luminosity distance, $\dl$, as a function of the redshift, $z$, are:
\begin{equation}
	\dc(z, \lc) = \int_0^z\diff{z^\prime} \frac{c}{H(z^\prime, \lc)}
	\label{eq:dc}
\end{equation}
and
\begin{equation}
	\dl(z, \lc) = (1+z)\dc(z, \lc),
	\label{eq:dl}
\end{equation}
where $c$ denotes the speed of the light, $\lc = \{\ho, \Om\}$ and
\begin{equation}
	H(z, \lc)=\ho \sqrt{\Om (1+z)^3 + 1 - \Om}.
	\label{eq:Hz}
\end{equation}
We use the fixed value of the fractional matter density in the \ac{MDA}, $\Om = 0.3$.
Considering
\begin{equation}
	\frac{\difff V_\text{c}}{\difff z} = \frac{\difff}{\difff z} \frac{4\pi}{3}\dc^3
	= 4\pi \dc^2 \frac{\difff \dc}{\difff z}
	= 4\pi \dc(z, \lc)^2 \frac{c}{H(z, \lc)},
\end{equation}
we obtain
\begin{equation}
	p(z|\hsignal, \lc) \propto\frac{1}{1+z} \frac{\dc(z, \lc)^2}{H(z, \lc)}.
	\label{eq:mda_pz}
\end{equation}

Putting \eqref{eq:conversion_Mcs_Mcd}, \eqref{eq:ropt}, \eqref{eq:mda_mass_model}, \eqref{eq:dc}, \eqref{eq:dl}, \eqref{eq:Hz} and \eqref{eq:mda_pz} into \eqref{eq:introducevTheta}, we obtain the observed \ac{SNR} distribution model for a given instrument.
The \ac{SNR} distribution model shown by the blue lines in \figref{fig:expected_snr_dist} assumes the sensitivity of the LIGO Hanford detector during the period from GPS time \mdagpstime, and it is employed in the \acp{MDA}.
For comparison, the gray lines and the orange lines in \figref{fig:expected_snr_dist} show the \ac{SNR} distribution models assuming the Advanced LIGO design sensitivity \cite{LIGO-T1800044} and the Cosmic Explorer design sensitivity \cite{LIGO-P1600143}, respectively.
The dependence of the distribution shape on the value of $\ho$ tends to be more pronounced in the low-\ac{SNR} region, and it can be seen that the better the assumed detector sensitivity---enabling distant \ac{CBC} to be observed with higher \ac{SNR}---the stronger the $\ho$-dependence of the distribution shape becomes.

By applying the importance sampling procedure described in \algref{alg:noise_signal_marg}, we obtained the distribution of the ranking statistic conditioned on a given \ac{SNR}, corresponding to the first term of \eqref{eq:signalmodel_reparametorization2} (shown in \figref{subfig:signalmodel_given_snr}).
The importance sampling was performed on the Kambai cluster at \ac{RESCEU}, The University of Tokyo.
Each node of the cluster is equipped with an Intel Xeon E5-2630 v4 CPU (20 threads) and 32 GB of memory.
The main computation consisted of 60 independent jobs with comparable computational cost.
The total wall time amounted to 637.9 hours (equivalent to 26.6 node-days) and the total CPU time to 3695.7 hours (user 3681.8 h, system 13.9 h).
The average peak memory usage per job was 31.7 GB, with a maximum of 39.4 GB.
In this study, for simplicity---as described in the first paragraph of \secref{subsubsec:combined_snr_pdf}---we impose as part of our detection criteria that all candidates included in our candidate list must be coincident across all detectors in the network (specifically, both the LIGO Hanford (H1) and LIGO Livingston (L1) detectors in our \acp{MDA}).
Accordingly, in \algref{alg:noise_signal_marg}, if a sample drawn from the proposal distribution corresponds to the set of instruments $\vec{O}$ other than $\{\text{H1, L1}\}$, such as $\vec{O} = \{\text{H1}\}$ or $\vec{O} = \{\text{L1}\}$, the signal model histogram $h_\text{signal}$ is not updated and the loop proceed to the next iteration.
Since the primary goal of this work is to validate the proposed framework, we did not invest significant effort in optimizing this part of the implementation.
As a result, the \acp{PDF} are likely to have converged less efficiently than what could be achieved with a more optimized implementation, given the computational resources expended.

Furthermore, using the single-detector \ac{SNR} distribution model shown by blue lines in \figref{fig:expected_snr_dist} together with the method described in \secref{subsubsec:combined_snr_pdf}, we constructed a two-detector combined \ac{SNR} distribution model (assuming the Hanford and Livingston detectors with the identical sensitivity).
By integrating these results thorough \eqref{eq:signalmodel_reparametorization2}, we obtained the single-candidate likelihood model under the signal hypothesis for different $\ho$, as shown in \figref{subfig:signalmodel_given_H0}.
As a reference, the original GstLAL signal model is plotted in orange, allowing a visual assessment of how the replacement of the spatial distribution model---described in \secref{subsec:rankcosmo}---affects the resulting distribution of the detection statistic.
As mentioned above, the LIGO detectors during the S5 run were not sensitive to very distant regions of the universe.
Consequently, the $\ho$-dependence of the model of the ranking statistic \ac{PDF} under the signal hypothesis remains limited.
We refer to the $\ho$-dependent distribution model of the detection statistic for astrophysical, signal-originating candidates, constructed through the process described so far, as the \ac{MDA} signal model.

\subsection{Data generation}\label{subsec:mda_data_generation}

We have used the inverse transform sampling to generate our synthetic dataset.
The noise candidates are drawn from the GstLAL noise model, and the signal candidates drawn from the combined \ac{SNR} distribution model.
Noise candidates are drawn directly from the \ac{PDF} model of the ranking statistic, subject to the requirement $-28\leq \ln\lr \leq 150$, and the data-generation process ends there.
For signal candidates, we first draw two-detector combined \ac{SNR} values from the \ac{SNR} distribution model constructed using S5 sensitivity, which corresponds to the injection value of $\ho$, subject to the requirement $7\leq\ac{SNR}\leq100$.
This lower \ac{SNR} threshold is imposed because the GstLAL software assigns a ranking statistic of negative infinity to triggers with network \ac{SNR} below 7.
Given the \ac{SNR} of each candidate, we then draw ranking statistic values from the conditional distribution, $p(x|\rcomb, \hsignal, \lambda)$, shown in \figref{subfig:signalmodel_given_snr}.
For also signal-originating candidates, we include only those satisfying $-28\leq \ln\lr\leq 150$ in the dataset.
This selection corresponds to the detection criteria, $\Delta$, introduced in the formalism section, \secref{sec:method}.
While in the section $\Delta$ was defined solely by a lower threshold on the detection statistic, here we also impose an upper threshold for practical convenience in normalizing probability distribution.
This additional restriction has no impact on the formalism itself.

By applying \eqref{eq:far} together with the actual counts of candidates observed between GPS times \mdagpstime, we obtain the \ac{IFAR} corresponding to the lower and upper detection thresholds, $\ln\lr=-28$ and $\ln\lr=150$, as 42 seconds and $1.4\times10^{72}$ years, respectively.
This demonstrates that our method is capable of handling catalogues that include candidates spanning a very broad range of statistical significance.

Although it is possible to generate signal candidates directly from the \ac{PDF} of the ranking statistic (the \ac{MDA} signal model), we chose not to do so, as this would obscure the connection to quantities such as \ac{SNR} that readers may find more intuitive.
A dataset based on component masses and redshifts would be even closer to astrophysical reality, but given the scope of this work we adopted a simpler, more tractable approach.
\figref{fig:comp_sample_originalpdf} shows the comparison between the example of the distribution of synthetic candidates and the modeled distributions.

In each mock universe, 10,000 candidates were generated.
The numbers of signal-originating and noise-originating candidates were determined so as to be consistent with the value of $\etabar$ assigned to each mock universe.

\subsection{Fidelity test of the signal fraction point estimate}\label{subsec:fraction_estimate_fidelity}
To further assess the fidelity of the signal fraction point estimation method described in \secref{subsec:fraction_estimate}, we conducted a dedicated validation test using a separate set of mock universes, distinct from the 1,386 mock universes used in the main \acp{MDA}.
This validation set consists of 1,500 mock universes, with injected values of $\etabar$ uniformly distributed between 0 and 1 in steps of $\Delta\etabar=2/1500$, while the Hubble constant $\ho$ is fixed at \SI{70}{\kilo\metre\per\second\per\mega\parsec} for all.
The data generation procedure follows the same approach as described in \secref{subsec:mda_data_generation}.
For universes with injected $\etabar<0.5$, we use a common list of 150,000 noise-origin candidates (ranking statistics), while only the signal-origin candidates vary across universes; the total number of candidates is not preserved.
Conversely, for injected $\etabar>0.5$, a common list of 150,000 signal-origin candidates is used, and only the noise-origin candidates differ between universes.
Again, the total number of candidates is not preserved.

\begin{figure*}
	\resizebox{0.86\linewidth}{!}{\includegraphics{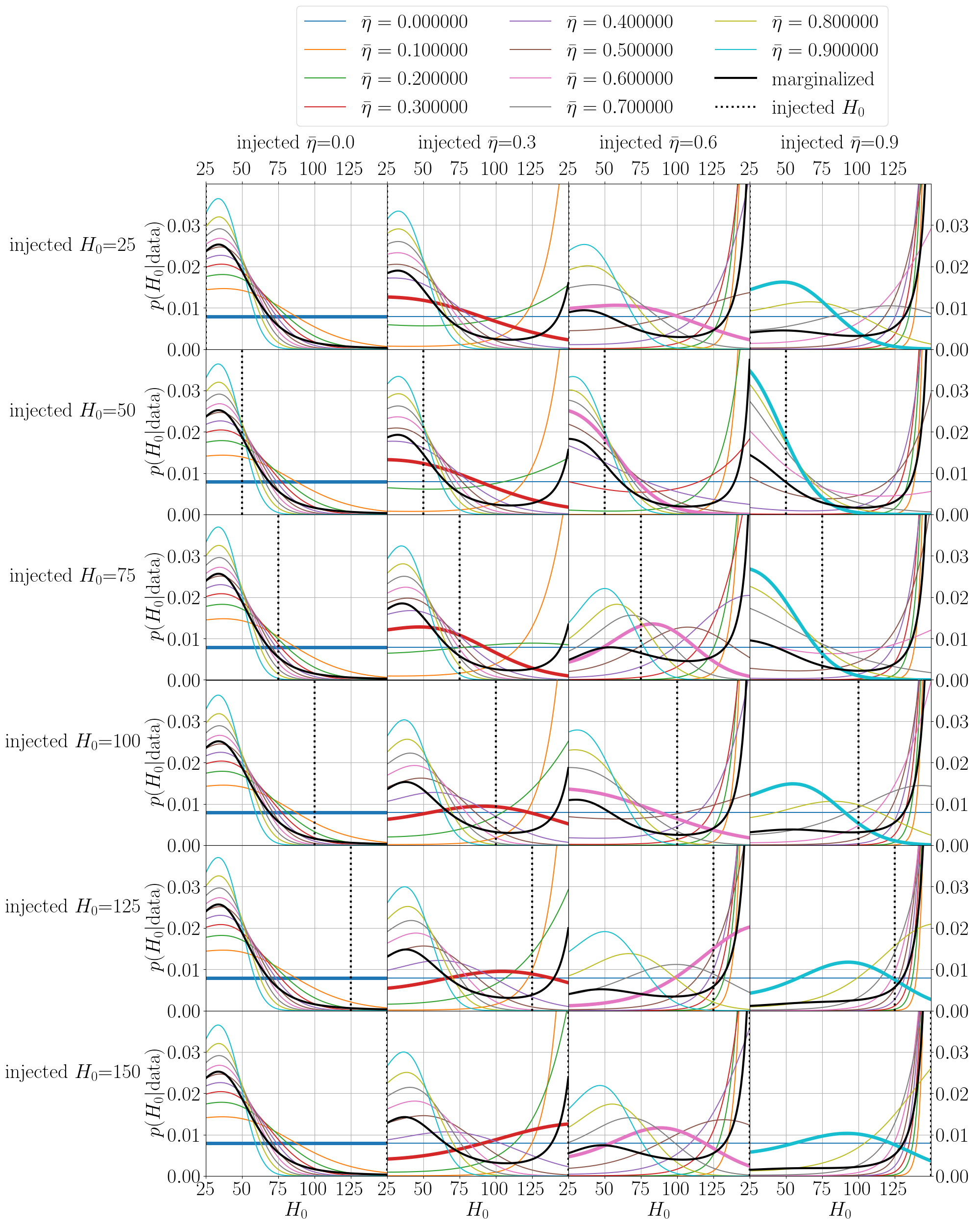}}
	\caption{Posteriors marginalized over the signal fraction, $\etabar$, from \acp{MDA} with different injected values of the Hubble constant $\ho$ and the signal fraction $\etabar$.
	Each row corresponds to a group of mock universes with different injected values of $\ho$, while each column corresponds to those with different injected values of $\etabar$.
In each panel, the black solid line shows the posterior of $\ho$ marginalized over $\etabar$, whereas the colored lines shows the posteriors of $\ho$ calculated under different assumed values of $\etabar$.
The curve corresponding to the true (injected) value of $\etabar$ for each mock universe is drawn with a thicker line, and the black dotted line indicates the true (injected) value of $\ho$.
When $\etabar=0$ is assumed in the evaluation of the posteriors (in the case of blue lines in each panel), no constraint on $\ho$ is obtained.
The marginalized posteriors show little sensitivity to the injected $\ho$ values, compared to their dependence on the injected $\etabar$ values.
In contrast, the posteriors obtained using the true $\etabar$ tend to track the injected $\ho$ values reasonably well.
}
	\label{fig:marginalized_posteriors}
\end{figure*}

\subsection{Results and interpretations}\label{subsec:mda_results}
\subsubsection{Results from joint estimate of \texorpdfstring{$\ho$}{ho} and \texorpdfstring{$\etabar$}{etabar}}\label{subsubsec:joint_estimate_results}
We performed a joint estimation of $\ho$ and $\etabar$ for each of the mock universes corresponding to 126 different values of $\ho$ and 11 different values of $\etabar$, using the synthetic observational data generated in \secref{subsec:mda_data_generation}.
\figref{fig:marginalized_posteriors} presents 24 of the resulting 1,386 posteriors for $\ho$.
Each row corresponds to a group of mock universes with different injected values of $\ho$, while each column corresponds to those with different injected values of $\etabar$.
In each panel, the black solid line shows the posterior of $\ho$ marginalized over $\etabar$, whereas the colored lines shows the posteriors of $\ho$ calculated under different assumed values of $\etabar$.
The curve corresponding to the true (injected) value of $\etabar$ for each mock universe is drawn with a thicker line, and the black dotted line indicates the true (injected) value of $\ho$.
Although only ten colored lines are shown, corresponding to $\etabar$ values from 0 to 0.9 in steps of 0.1, in each mock universe we actually computed the $\ho$ posterior for 1,000 $\etabar$ values between 0 and 1, which were subsequently marginalized.
As expected, when $\etabar=0$ is assumed in the evaluation of the posteriors (in the case of blue lines in each panel), all the candidates in our datasets are interpreted as noise, and thus no constraint on $\ho$ is obtained.
From a global perspective, the marginalized posteriors (black solid lines) show little sensitivity to the injected $\ho$ values, compared to their dependence on the injected $\etabar$ values.
This is consistent with the discussion in the previous subsection and with \figref{subfig:signalmodel_given_H0}: because the sensitivity of the initial LIGO detector in the S5 era was not sufficient to probe distant sources, the shape of the ranking statistic distribution is only weakly dependent on $\ho$.
In contrast, the posteriors obtained using the true $\etabar$ tend to track the injected $\ho$ values reasonably well.
This behavior motivates the use of a point estimate of $\etabar$ based on $\pastro$, as introduced in \secref{subsec:fraction_estimate}.

\begin{figure}
	\resizebox{0.9\linewidth}{!}{\includegraphics{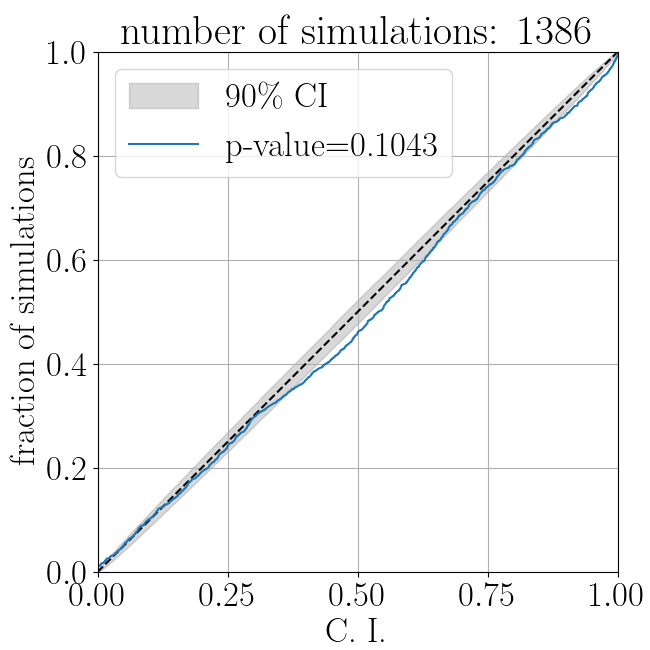}}
	\caption{Probability-probability plot for the joint estimate of $\ho$ and $\etabar$, derived from the analyses of 1,386 mock universes, corresponding to 126 different values of the Hubble constant $\ho$ and 11 different values of the signal fraction $\etabar$.
	The shaded region indicates the 90\% confidence interval expected from the binomial distribution and the p-value shown in the legend is obtained from \ac{KS} test \cite{scipy}.
	While some deviation from the confidence band is observed, no significant inconsistency is found between the recovered posteriors and the injected values.
	}
	\label{fig:pp_plot_joint}
\end{figure}

\figref{fig:pp_plot_joint} shows the \ac{p-p} plot obtained from the analyses of 1,386 mock universes.
The blue line shows the fraction of mock universes whose injected $\ho$ values fall within a given credible interval (C.I.) of the inferred posterior.
The credible intervals are calculated in a one-sided (left-tailed) manner.
The shaded region indicates the 90\% confidence interval expected from the binomial distribution.To quantify deviations from uniformity, we performed a \ac{KS} test using SciPy's \texttt{kstest} function \cite{scipy}, yielding a $p$-value of $\sim 0.1$.
While some deviation from the confidence band is observed, no significant inconsistency is found between the recovered posteriors and the injected values.

\subsubsection{Results of fidelity test of \texorpdfstring{$\etabar$}{etabar} point estimate}\label{subsubsec:result_fiedlity}
\figref{fig:fraction_estimate_fidelity} presents the results of the signal fraction fidelity test described in \secref{subsec:fraction_estimate_fidelity}.
In the main panel, the horizontal axis shows the true (injected) signal fraction $\etabar_0$, while the vertical axis shows the signal fraction estimated using our point estimate method based on $\pastro$.
The black dashed line represents the ideal relation in which the estimated value exactly matches the true value.
The thick light gray solid line shows the relationship between $\etabar_0$ and the estimated value $\etabar_1^\prime$, obtained as a simple sum of $\pastro$ values divided by the total number of candidates as described in \eqref{eq:pastro_sum}.
While this line is linear like the ideal case, its slope differs, and it intersects the ideal diagonal near $\etabar_0=0.5$.
The upper subpanel shows the corresponding relative error $|\etabar_1^\prime - \etabar_0| / \etabar_0$ as a light gray solid line.
The error decreases near $\etabar_0=0.5$ and increases toward the edges of the plot; in particular, it exceeds 100\% for small values of $\etabar_0$.
The dark gray dotted line in the main panel indicates the corrected estimate $\etabar_0^\prime$ obtained using \eqref{eq:def_gammabar_0_prime}, which closely follows the ideal relation.
The upper subpanel shows the corresponding relative error $|\etabar_0^\prime - \etabar_0| / \etabar_0$ as a dark gray dotted line, revealing that the error decreases as $\etabar_0$ increases.
For $\etabar_0\gtrsim0.03$, the corrected estimate achieves better than 5\% accuracy, and the error drops below 1\% near $\etabar_0=1$.
The blue solid line in the main panel is a least-squares fit to the linear relationship between $\etabar_0$ and $\etabar_1^\prime$, with the best-fit parameters shown in the legend.
The orange dashed line represents the linear prediction based on \eqref{eq:true_estimate_relation}, and its coefficients are also indicated in the legend.
This predicted relationship agrees well with the observed trend (light gray solid line and blue fit).
This agreement, in turn, enables the accurate correction illustrated by the dark gray dotted line.

	As briefly mentioned in the end of \secref{subsec:fraction_estimate}, from the conceptual point of view of the point estimation, it is not necessary to use $\pastro$.
	In principle, one could construct an analogous estimator by replacing $\pastro$, $p(\hsignal|x, \lambda, \Delta)$ in \eqref{eq:pastro_sum}, \eqref{eq:pastro_exp_signal} and \eqref{eq:pastro_exp_noise} with any function of the detection statistic, such as the detection statistic $x$ itself.
	However, when we performed the same fidelity test using this replacement, the relative error showed a similar monotonic decrease with $\etabar_0$ but remained larger overall: the error did not fall below 1\% even near $\etabar_0=1$, and the true signal fraction needed for the error to drop below 5\% was $\etabar_0\gtrsim 0.1$.
	These results empirically justify our choice of employing $\pastro$ in the estimator.

\subsubsection{Results using point estimate of signal fraction}\label{subsubsec:results_using_point_estimate}
\begin{figure}
	\resizebox{\linewidth}{!}{\includegraphics{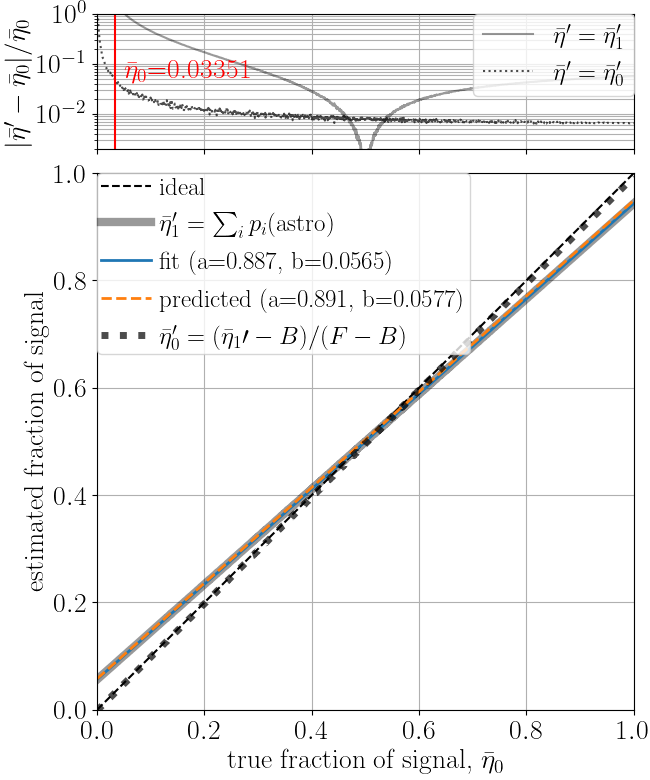}}
	\caption{True signal fraction vs.~estimated signal fraction.
	In the bottom main panel, the horizontal axis shows the injected signal fraction $\etabar_0$, while the vertical axis shows the signal fraction estimated using our point estimate method based on $\pastro$.
	The upper subpanel shows the relative estimation error $|\etabar^\prime - \etabar_0| / \etabar_0$.
	For the estimator defined in \eqref{eq:pastro_sum}, $\etabar_1^\prime$, the error decreases near $\etabar_0=0.5$ and increases toward the edges of the plot; it exceeds 100\% for small values of $\etabar_0$.
	For the estimator defined in \eqref{eq:def_gammabar_0_prime}, $\etabar_0^\prime$, the error decreases as $\etabar_0$ increases.
	For $\etabar_0\gtrsim0.03$, $\etabar_0^\prime$ achieves better than 5\% accuracy, and the error drops below 1\% near $\etabar_0=1$.
	}
	\label{fig:fraction_estimate_fidelity}
\end{figure}

\begin{figure*}
	\resizebox{0.9\linewidth}{!}{\includegraphics{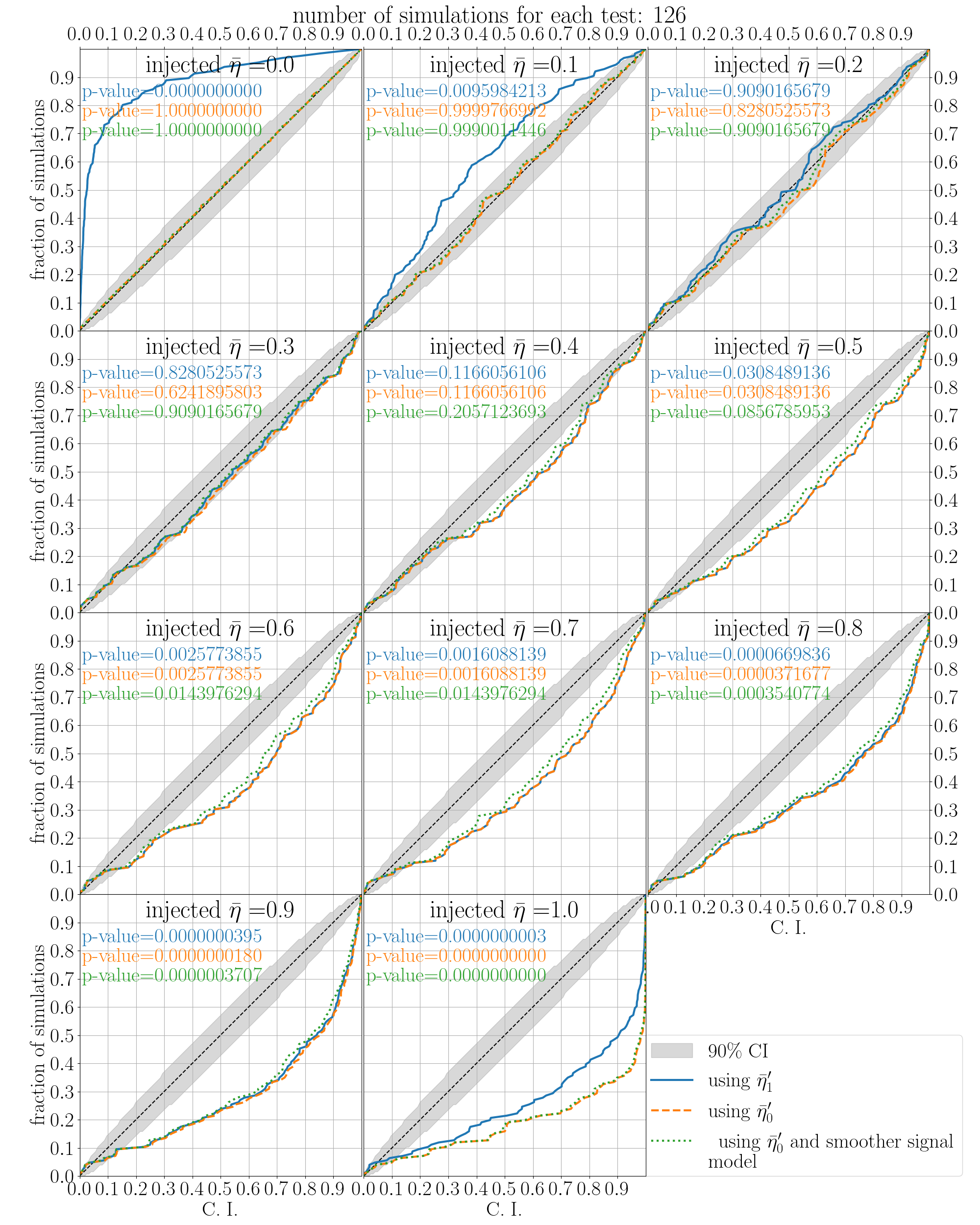}}
	\caption{Probability-probability plots for the inferences using the point estimate of the signal fraction $\etabar$ with different injection values of $\etabar$.
	In each of 11 panels, the gray shaded region represents the 90\% frequentist confidence band, same as in \figref{fig:pp_plot_joint}.
	The blue solid and orange dashed lines indicate the results of the \ac{p-p} tests assuming $\etabar^\prime = \etabar_1^\prime$ and $\etabar^\prime = \etabar_0^\prime$, respectively.
	The green dotted lines represents the \ac{p-p} plot derived from the smoother signal model, calculated via importance sampling using 4 times as many samples as the \ac{MDA} signal model shown in \figref{subfig:signalmodel_given_H0}.
	For injected values of $\bar{\eta} \lesssim 0.2$, we find that the Hubble constant can be inferred without noticeable bias when using a reliable signal fraction estimator.
	In contrast, for $\bar{\eta} \gtrsim 0.3$, the impact of underestimation of the hierarchical likelihood at specific $\ho$ values, induced by the $\ho$-dependent fluctuation of the \ac{MDA} signal model, becomes significant.
	}
	\label{fig:pp_plots_diff_fraction}
\end{figure*}

\figref{fig:pp_plots_diff_fraction} shows a set of 11 \ac{p-p} plots, each corresponding to a group of 126 mock universes sharing the same injected value of $\etabar$.
The analyses in each panel are based on the Dirac delta prior introduced in \eqref{eq:dirac_delta_prior}.
In general, any mismatch between the population distribution used to generate the simulations and the prior assumed in Bayesian inference appears as a bias in the resulting p–p plot.
However, as confirmed in \secref{subsubsec:result_fiedlity}, the difference between our estimate $\etabar^\prime$ and the injected value $\etabar_0$ is small, at least in the case of $\etabar^\prime = \etabar_0^\prime$, that we can approximate $\delta(\etabar - \etabar^\prime)\simeq\delta(\etabar - \etabar_0)$, and thereby construct \ac{p-p} plots for each value of the injected $\etabar$ (or $\etabar_0$).
In each of 11 panels, the gray shaded region represents the 90\% frequentist confidence band, same as in \figref{fig:pp_plot_joint}.
The blue solid and orange dashed lines indicate the results of the \ac{p-p} tests assuming $\etabar^\prime = \etabar_1^\prime$ and $\etabar^\prime = \etabar_0^\prime$, respectively.
The green dotted lines will be discussed later.
The results shown in the panels corresponding to small injected values of $\etabar$, specifically $\etabar\lesssim0.2$, behave as expected.
The orange dashed lines lie close to the diagonal, and the associated \ac{KS} test p-values (shown in the same color) are near unity.
In contrast, for injected $\etabar=0$ and 0.1, the blue solid lines deviate substantially from the diagonal due to large discrepancies between the estimate $\etabar_1^\prime$ and the injected value, resulting in small \ac{KS} p-values.
As anticipated from the results in \figref{fig:fraction_estimate_fidelity}, the blue curves become progressively better as the injected $\etabar$ approaches 0.5, eventually showing similar behavior to the orange dashed curves.
For the injected $\etabar\gtrsim 0.3$, both the blue solid and orange dashed lines begin to deviate noticeably from the diagonal.
Ideally, we would expect the orange dashed lines to remain close to the diagonal for all the injected values of $\etabar$ while the blue solid lines should show strong deviations for the injected $\etabar\simeq 0.9$ and 1---similar in magnitude to the deviations seen for the injected $\etabar=0$ and 0.1.
However, this expected pattern is not observed.
We hypothesize that the fact that both the blue and orange curves exhibit similarly strong biases in the mock universes with large injected $\etabar$ indicates the presence of a systematic error that has not been accounted for in our previous explanations.
This systematic effect appears to dominate over the differences between $\etabar_0^\prime$ and $\etabar_1^\prime$.
We attribute its origin to the lack of sufficient convergence in our \ac{MDA} signal model constructed via importance sampling, which results in residual statistical fluctuations in the modeled signal distribution.

In our main \acp{MDA}, the same non-converged \ac{MDA} signal model---with residual statistical fluctuations---is used both in data generation and in the evaluation of the hierarchical likelihood.
However, since data generation is inherently stochastic, we expect that the impact of fluctuations in the \ac{MDA} signal model is relatively minor in this step: synthetic candidate lists drawn from the noisy signal model are likely to be similar to those drawn from a fully converged model.
In contrast, the hierarchical likelihood evaluation is more sensitive to such fluctuations, and we hypothesize that this asymmetry is responsible for the biases observed in many of the \ac{p-p} plots shown in \figref{fig:pp_plots_diff_fraction}.

To prepare for a mathematical explanation of the bias introduced in the evaluation of the hierarchical likelihood due to statistical fluctuations in the \ac{MDA} signal model, we begin by defining two detection-statistic probability density models.
The true, fully converged, fluctuation-free model is defined, following \eqref{eq:reduced_hl}, as:
\begin{equation}
	\ptrue(x|\ho, \etabar) = \etabar \ptrue(x|\hsignal, \ho) + (1 - \etabar)\ptrue(x|\hnoise).
\end{equation}
The fluctuation-affected model, constructed via importance sampling using a finite number of samples, is denoted as:
\begin{equation}
	\phat(x|\ho, \etabar) = \ptrue(x|\ho, \etabar) + \etabar\epsilon(x|\ho),
\end{equation}
where $\epsilon(x|\ho)$ characterizes the statistical fluctuation present primarily in the signal component (the \ac{MDA} signal model).
This modeling is motivated by the empirical observation (see \figref{subfig:lr_hist}) that the noise model exhibits much smaller fluctuations than the signal model.
Even if there are small residual fluctuations in the noise component, they are independent of $\ho$ and thus do not affect the relative likelihood between different $\ho$ values.
Therefore, we neglect the contribution of fluctuations from the noise model and model the total \ac{PDF} fluctuation using only the signal-side variation scaled by $\etabar$.
In our practical implementation, we evaluate the logarithmic form of the hierarchical likelihood.
For a fluctuation-free model, this takes the form:
\begin{equation}
	\ln \ptrue(D|\ho, \etabar) = \sum_{i=1}^{\Ntot}\ln\ptrue(x_i|\ho, \etabar).
\end{equation}
In contrast, when using the fluctuation-affected model, the likelihood becomes:
\begin{align}
\ln \phat(D|\ho, \etabar)
	=\sum_{i=1}^{\Ntot}[\ln\ptrue(x_i|\ho, \etabar) + \etabar\epsilon(x_i|\ho)]\\
	= \ln \ptrue(D|\ho, \etabar) + \sum_{i=1}^{\Ntot}\ln\left[1 + \frac{\etabar\epsilon(x_i|\ho)}{\ptrue(x_i|\ho, \etabar)}\right].
\end{align}
Denoting the ensemble average over the realization of the importance-sampled model as $\average{\cdot}$, and assuming $\average{\frac{\etabar\epsilon(x|\ho)}{\ptrue(x|\ho, \etabar)}}=0$, from the Maclaurin expansion to the leading order, we obtain
\begin{equation}
	\average{\ln\left[1 + \frac{\etabar\epsilon(x_i|\ho)}{\ptrue(x_i|\ho, \etabar)}\right]}\simeq -\frac{1}{2}\average{\left[\frac{\etabar\epsilon(x|\ho)}{\ptrue(x|\ho, \etabar)}\right]^2}.
\end{equation}
Consequently, the expected log hierarchical likelihood evaluated with the fluctuation-affected model is always smaller than the true value:
\begin{align}
	&\average{\ln \phat(D|\ho, \etabar)}\nonumber\\
	&\simeq\ln \ptrue(D|\ho, \etabar) - \frac{1}{2}\sum_{i=1}^{\Ntot} \average{\left[\frac{\etabar\epsilon(x_i|\ho)}{\ptrue(x_i|\ho, \etabar)}\right]^2}
	\label{eq:expected_deviation_from_ptrue}\\
	&\leq \ln \ptrue(D|\ho, \etabar)\nonumber.
\end{align}
Therefore, if the magnitude of fluctuation varies with $\ho$, this bias can lead to spurious suppression of the hierarchical likelihood at particular values of $\ho$, independent of their proximity to the injected truth.
It can also be found that the systematic underestimation of the hierarchical likelihood becomes more severe as either $\etabar$ increases, or $\Ntot$ increases, since the second term of \eqref{eq:expected_deviation_from_ptrue} increases monotonically with both variables.

\begin{figure}
	\resizebox{0.9\linewidth}{!}{\includegraphics{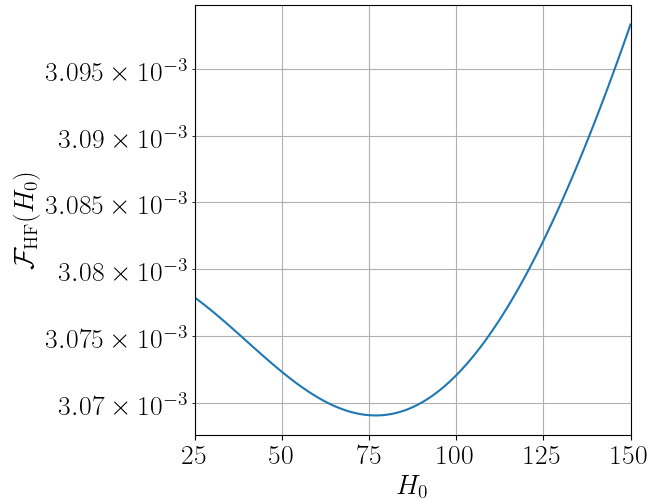}}
	\caption{The fraction of power contained in fluctuations with wavelengths shorter than the wavelength $\Delta x = \Delta\ln\lr\leq 5$ of the detection statistic distribution, $p(x|\hsignal, \ho)$, for each value of $\ho$.
	It is observed that the signal models corresponding to larger values of $\ho$ have a greater fraction of high-frequency oscillation power.
	}
	\label{fig:smoothness}
\end{figure}

To quantitatively evaluate the $\ho$-dependence of the magnitude of statistical fluctuations in our \ac{MDA} signal model, we computed the fraction of power contained in fluctuations with wavelengths shorter than the wavelength $\Delta x = \Delta\ln\lr\leq 5$ of the detection statistic distribution, $p(x|\hsignal, \ho)$, as defined below, for each value of $\ho$:
\begin{equation}
	\frachf(\ho) \equiv \frac{\int_{f\geq 1/\Delta x}\diff{f}|\tilde{p}(f|\ho)|^2}{\int_{f>0}\diff{f}|\tilde{p}(f|\ho)|^2},
\end{equation}
where $\tilde{p}(f|\ho)$ is the Fourier transform of $p(x|\hsignal, \ho) - \bar{p}(\ho)$, and $\bar{p}(\ho)$ is the DC (zero-frequency) component of $p(x|\hsignal, \ho)$.
\figref{fig:smoothness} shows the dependence of $\frachf$ on $\ho$, computed for our \ac{MDA} signal model.
It is observed that the signal models corresponding to larger values of $\ho$ have a greater fraction of high-frequency oscillation power.

\begin{figure*}
	\resizebox{0.86\linewidth}{!}{\includegraphics{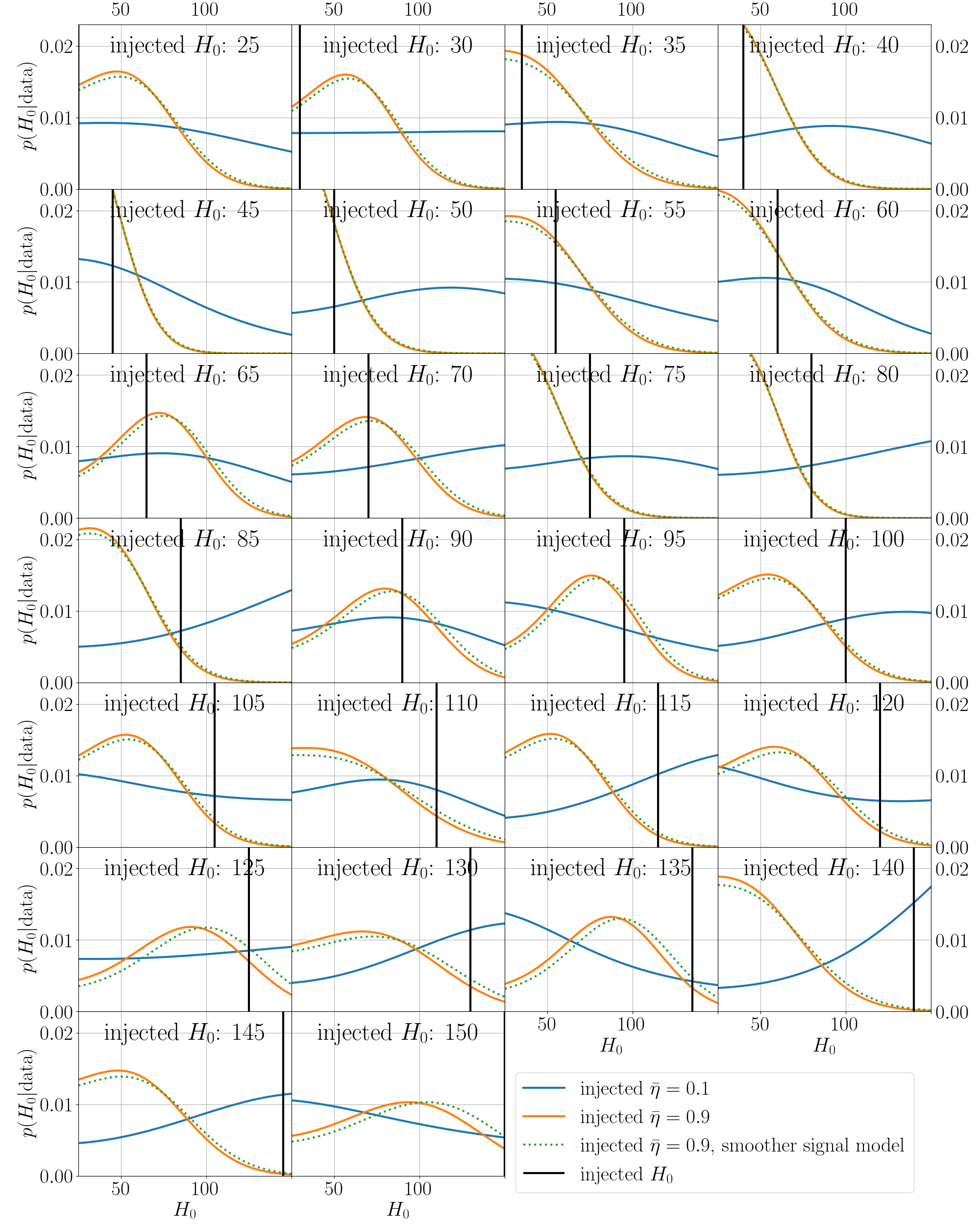}}
	\caption{Comparison of the posteriors obtained using the point estimate of the signal fraction, between the different injection value of the signal fraction, $\etabar=0.1$ and $0.9$.
	 The posteriors corresponding to the injected $\etabar=0.9$, shown by orange lines, consistently exhibit suppression around large $\ho$ values, regardless of the injected value of $\ho$.
	This behavior is consistent with the discussion leading up to \eqref{eq:expected_deviation_from_ptrue} and with the trend in the magnitude of model fluctuations shown in \figref{fig:smoothness}.
	In contrast, the blue curves corresponding to injected $\etabar=0.1$ show no such suppression, supporting the interpretation from \eqref{eq:expected_deviation_from_ptrue} that the systematic underestimation of the hierarchical likelihood becomes more severe as $\etabar$ increases.
	The green dotted lines also show the posterior for the mock universe with the injected $\etabar=0.9$, and the analyzed datasets are same as those of the orange lines.
	However, the posteriors shown by the green dotted lines obtained from the hierarchical likelihood evaluation using the smoother signal model than the original \ac{MDA} signal model.
	When using a smoother signal model for the hierarchical likelihood evaluation, the suppression at large $\ho$ becomes relatively less pronounced.
	}
	\label{fig:posterior_gamma_comp}
\end{figure*}

\figref{fig:posterior_gamma_comp} shows a subset of the posteriors obtained from the \acp{MDA} assuming a Dirac delta prior.
Focusing on the orange curves, which correspond to the case of injected $\etabar=0.9$, we observe that the posteriors consistently exhibit suppression around large $\ho$ values, regardless of the injected value of $\ho$.
This behavior is consistent with the discussion leading up to \eqref{eq:expected_deviation_from_ptrue} and with the trend in the magnitude of model fluctuations shown in \figref{fig:smoothness}.
In contrast, the blue curves corresponding to injected $\etabar=0.1$ show no such suppression, supporting the interpretation from \eqref{eq:expected_deviation_from_ptrue} that the systematic underestimation of the hierarchical likelihood becomes more severe as $\etabar$ increases.
Furthermore, the suppression of the hierarchical likelihood at large $\ho$ implies that the resulting left-sided Bayesian credible intervals may be underestimated.
This is consistent with the observation that the \ac{p-p} plot in \figref{fig:pp_plot_joint} and those for mock universe groups with $\etabar\gtrsim 0.3$ in \figref{fig:pp_plots_diff_fraction} exhibit a systematic sagging below the diagonal.
The green dotted line in \figref{fig:posterior_gamma_comp} shows the posterior distribution obtained by analyzing mock data with injected $\etabar=0.9$ using a different signal model from our original \ac{MDA} signal model constructed in \secref{subsec:mda_pop_model}.
The observed data (i.e., the \ac{GW} candidate list) are exactly the same as those used for the orange solid lines.
The difference lies in the signal model, which is slightly smoother because it was constructed via importance sampling using 4 times as many samples as the original model.
Although the difference is subtle, the green dotted posteriors are consistently shifted toward larger $\ho$ values compared to the orange curves, which is consistent with our previous discussions.
In addition, the green dotted lines in \figref{fig:pp_plots_diff_fraction} represent the \ac{p-p} plot derived from this smoother signal model.
While the improvement is marginal, it lies slightly closer to the diagonal than the orange line.

\section{Conclusion}\label{sec:conclusion}
We have presented a Bayesian hierarchical framework for inferring population parameters from unedited compact object merger catalogues---i.e., candidates lists produced directly by a search pipeline, without the application of additional selection thresholds (e.g., $\pastro\geq 0.5$) as typically used in GWTC catalogues \cite{GWTC-1, GWTC-2, GWTC-2.1, GWTC-3, GWTC-4.0}.
Crucially, this framework does not require per-candidate parameter estimation and instead operates directly on the detection statistic values assigned to each candidate.
Our approach is a practical variant of the general hierarchical inference framework developed by \cite{FGMC}.

To validate the method, we applied it to a large number of mock datasets and focused on the joint inference of the Hubble constant $\ho$ and the signal fraction $\etabar$, relying on the data products generated in the S5S6 reanalysis \cite{s5s6_2021_reanalysis}.
	The $\ho$-dependence of the \ac{MDA} signal model remains limited, since the LIGO detectors during the S5 run were not sensitive to very distant regions of the universe.
	Consequently, our \acp{MDA} mostly unable to recover the injected $\ho$.
	However, the probability-probability test demonstrated that the recovered posteriors are statistically consistent with the injected values: the obtained \ac{p-p} plot lies close to the diagonal, with a p-value of $\sim0.1$ obtained from the KS test \cite{scipy}.
	While a slight downward deviation from the diagonal is observed, it is consistent with the bias expected from the incomplete convergence of the \ac{MDA} signal model discussed in \secref{subsubsec:results_using_point_estimate}.
	Although some convergence issue of the signal model remains, the overall result indicates that our inference framework itself is essentially free from bias, as evidenced by the \ac{p-p} test.
	This outcome also substantiates our initial expectation that, by using a dataset consisting of the detection statistics themselves, which already reflect the search threshold, the framework can naturally incorporate the selection imposed by the search without requiring explicit injection-based sensitivity estimates.

As a proof of concept, our \acp{MDA} were designed to highlight the conceptual structure of the framework rather than to achieve a fully realistic cosmological inference.
To reduce computational cost and complexity, we assumed an idealized situation in which the true intrinsic parameter distribution of compact binaries is already perfectly known and is consistently used as the population model in the detection pipeline.
Accordingly, our mock universes were constructed to follow the same mass model as adopted in the S5S6 reanalysis \cite{s5s6_2021_reanalysis}, namely a log-uniform distribution in the detector-frame mass.
Our simplified assumption is not realistic, and furthermore, previous studies have highlighted the importance of jointly estimating the mass distribution and cosmological parameters to avoid biases \cite{gwtc-3_cosmo, gwtc-4_cosmo}, as well as the potential improvement in precision when using more structured mass models featuring, for example, mass gaps or peaks \cite{Taylor_2012, Farr_2019,Mapelli_2020, Ezquiaga_2022, Mali_2024}.
As discussed in \secref{sec:gstlal_lr}, however, the detection statistic generated by a search pipeline is computed under a specific population model, and its distribution changes when the underlying population changes.
In our implementation, we modified the GstLAL signal model and used it as the single-candidate likelihood under the signal hypothesis.
If one were to jointly estimate cosmological and other population parameters, this modification could not be limited to replacing the SNR distribution model as done in \eqref{eq:signalmodel_reparametorization2}, but would require a full modification of the signal model as described in \eqref{eq:signalmodel_reparametorization}.
Even in our simplified \acp{MDA}, incomplete convergence of the signal model posed a problem, and such a full modification would further increase the dimensionality of the \ac{PDF} estimated via importance-weighted sampling, making convergence more computationally demanding.
Understanding how well the signal model needs to be converged in order to ensure unbiased inference will be an important topic for future study, particularly in the context of simultaneous estimation of multiple population parameters.
Our results indicate that the bias arising from incomplete convergence tends to increase with the signal fraction and the total number of candidates, suggesting that achieving sufficient convergence will become increasingly critical for precision inference of cosmological or astrophysical population parameters in future analyses.

As a possible direction for addressing this convergence issue, we note that the importance-weighted sampling algorithm used to construct the signal model---adapted from the GstLAL programs \cite{gstlal}---was originally optimized for building the noise model rather than the signal model.
As can be seen in \figref{subfig:signalmodel_given_snr}, the model looks smooth and well converged in the noise-like region with small $\ln\lr$, while in the more signal-like region with large $\ln\lr$, it exhibits noticeable fluctuations.
This suggests that the current algorithm efficiently extracts noise-like samples but is less effective for sampling signal-like events.
Developing algorithms that can efficiently address this issue will be essential for future analyses that aim to extract maximal information from large \ac{GW} datasets through joint estimation of population and cosmological parameters.

Although the joint inference of signal fraction and population parameters has been explored in earlier works \cite{FGMC, Heinzel_2023}, our analyses emphasize that $\etabar$ can be separately estimated from the observational data using a method based on the average of $\pastro$.
Specifically, we showed that a point estimate of $\etabar$, derived from the linearly transformed average of $\pastro$, closely matches the injected value.
When this estimate is used to construct a Dirac delta prior for the signal fraction, the resulting population inferences---in our \acp{MDA}, constraints on $\ho$---becomes more informative than those obtained from joint inference.
We presented both the results of joint inference and the results obtained by fixing $\etabar$ to the point estimate.
While the joint posteriors showed no significant bias, the Dirac-prior analyses yielded noticeably tighter constraints, confirming that leveraging a reliable signal fraction estimate enhances the informativeness of population-level inference.

While one of the main strengths of our framework lies in not requiring per-candidate parameter estimation---allowing us to handle catalogues consisting of a large number of candidates---this advantage inherently involves a trade-off between the breadth of information gained from many candidates and the depth of information obtainable from the detailed characterization of each candidate.
It will therefore be an interesting subject for future studies to investigate, for detectors of various sensitivities and observation durations, under what conditions this method can achieve constraints on population parameters, including $\ho$, that are comparable to those obtained with conventional population-inference method.

\acknowledgments
The authors are grateful for computational resources provided by the LIGO Laboratory and supported by National Science Foundation grants PHY-0757058 and PHY-0823459.
This research has made use of LALSuite software \cite{lalsuite}.
RH was supported by JST SPRING (JPMJSP2108).
RH and KC were supported by JSPS KAKENHI grants JP18H03698 and JP23H04893.
HF acknowledges support from the NSERC Alliance International Collaboration program.

\bibliography{article}

@article{FGMC,
   title={Counting and confusion: Bayesian rate estimation with multiple populations},
   volume={91},
   ISSN={1550-2368},
   url={http://dx.doi.org/10.1103/PhysRevD.91.023005},
   DOI={10.1103/physrevd.91.023005},
   number={2},
   journal={Phys. Rev. D},
   publisher={American Physical Society (APS)},
   author={Farr, Will M. and Gair, Jonathan R. and Mandel, Ilya and Cutler, Curt},
   year={2015},
   month=jan }

@article{Gaebel_2019,
    author = {Gaebel, Sebastian M and Veitch, John and Dent, Thomas and Farr, Will M},
    title = {{Digging the population of compact binary mergers out of the noise}},
    journal = {Mon. Not. R. Astron. Soc.},
    volume = {484},
    number = {3},
    pages = {4008-4023},
    year = {2019},
    month = {01},
    issn = {0035-8711},
    doi = {10.1093/mnras/stz225},
    url = {https://doi.org/10.1093/mnras/stz225},
    eprint = {https://academic.oup.com/mnras/article-pdf/484/3/4008/27738367/stz225.pdf},
}

@article{Shanika_2020,
  title = {Gravitational-wave inference in the catalog era: Evolving priors and marginal events},
  author = {Galaudage, Shanika and Talbot, Colm and Thrane, Eric},
  journal = {Phys. Rev. D},
  volume = {102},
  issue = {8},
  pages = {083026},
  numpages = {24},
  year = {2020},
  month = {Oct},
  publisher = {American Physical Society},
  doi = {10.1103/PhysRevD.102.083026},
  url = {https://link.aps.org/doi/10.1103/PhysRevD.102.083026}
}

@article{Roulet_2020,
  title = {Binary black hole mergers from LIGO/Virgo O1 and O2: Population inference combining confident and marginal events},
  author = {Roulet, Javier and Venumadhav, Tejaswi and Zackay, Barak and Dai, Liang and Zaldarriaga, Matias},
  journal = {Phys. Rev. D},
  volume = {102},
  issue = {12},
  pages = {123022},
  numpages = {24},
  year = {2020},
  month = {Dec},
  publisher = {American Physical Society},
  doi = {10.1103/PhysRevD.102.123022},
  url = {https://link.aps.org/doi/10.1103/PhysRevD.102.123022}
}

@inbook{Vitale_2021,
   title={Inferring the Properties of a Population of Compact Binaries in Presence of Selection Effects},
   ISBN={9789811547027},
   url={http://dx.doi.org/10.1007/978-981-15-4702-7_45-1},
   DOI={10.1007/978-981-15-4702-7_45-1},
   booktitle={Handbook of Gravitational Wave Astronomy},
   publisher={Springer Singapore},
   author={Vitale, Salvatore and Gerosa, Davide and Farr, Will M. and Taylor, Stephen R.},
   year={2021},
   pages={1–60}
}

@article{Heinzel_2023,
   title={Inferring the astrophysical population of gravitational wave sources in the presence of noise transients},
   volume={523},
   ISSN={1365-2966},
   url={http://dx.doi.org/10.1093/mnras/stad1823},
   DOI={10.1093/mnras/stad1823},
   number={4},
   journal={Mon. Not. R. Astron. Soc.},
   publisher={Oxford University Press (OUP)},
   author={Heinzel, Jack and Talbot, Colm and Ashton, Gregory and Vitale, Salvatore},
   year={2023},
   month=jun, pages={5972–5984}
}

@misc{Mehta_2025,
      title={Binary black hole population inference combining confident and marginal events from the $\tt{IAS\text{-}HM}$ search pipeline},
      author={Ajit Kumar Mehta and Digvijay Wadekar and Isha Anantpurkar and Javier Roulet and Tejaswi Venumadhav and Tousif Islam and Jonathan Mushkin and Barak Zackay and Matias Zaldarriaga},
      year={2025},
      eprint={2508.15350},
      archivePrefix={arXiv},
      primaryClass={gr-qc},
      url={https://arxiv.org/abs/2508.15350},
}

@article{gwtc-3_pop,
  title = {Population of Merging Compact Binaries Inferred Using Gravitational Waves through GWTC-3},
  author = {Abbott, R. and others},
  collaboration = {LIGO Scientific Collaboration, Virgo Collaboration, and KAGRA Collaboration},
  journal = {Phys. Rev. X},
  volume = {13},
  issue = {1},
  pages = {011048},
  numpages = {75},
  year = {2023},
  month = {Mar},
  publisher = {American Physical Society},
  doi = {10.1103/PhysRevX.13.011048},
  url = {https://link.aps.org/doi/10.1103/PhysRevX.13.011048}
}

@article{non-central-chi-square,
 ISSN = {00063444, 14643510},
 URL = {http://www.jstor.org/stable/2332542},
 author = {P. B. Patnaik},
 journal = {Biometrika},
 number = {1/2},
 pages = {202--232},
 publisher = {[Oxford University Press, Biometrika Trust]},
 title = {The Non-Central chi-squared- and F-Distribution and their Applications},
 urldate = {2025-09-19},
 volume = {36},
 year = {1949}
}

@article{scipy,
   title={SciPy 1.0: fundamental algorithms for scientific computing in Python},
   volume={17},
   ISSN={1548-7105},
   url={http://dx.doi.org/10.1038/s41592-019-0686-2},
   DOI={10.1038/s41592-019-0686-2},
   number={3},
   journal={Nat. Methods},
   publisher={Springer Science and Business Media LLC},
   author={Virtanen, Pauli and others},
   year={2020},
   month=feb,
   pages={261–272}
}

@article{Guglielmetti_2009,
   title={Background-source separation in astronomical images with Bayesian probability theory - I. The method},
   volume={396},
   ISSN={1365-2966},
   url={http://dx.doi.org/10.1111/j.1365-2966.2009.14739.x},
   DOI={10.1111/j.1365-2966.2009.14739.x},
   number={1},
   journal={Mon. Not. R. Astron. Soc.},
   publisher={Oxford University Press (OUP)},
   author={Guglielmetti, F. and Fischer, R. and Dose, V.},
   year={2009},
   month=jun, pages={165–190}
}

@article{BBHs_in_O1,
  title = {Binary Black Hole Mergers in the First Advanced LIGO Observing Run},
  author = {Abbott, B. P. and others},
  collaboration = {LIGO Scientific Collaboration and Virgo Collaboration},
  journal = {Phys. Rev. X},
  volume = {6},
  issue = {4},
  pages = {041015},
  numpages = {36},
  year = {2016},
  month = {Oct},
  publisher = {American Physical Society},
  doi = {10.1103/PhysRevX.6.041015},
  url = {https://link.aps.org/doi/10.1103/PhysRevX.6.041015}
}

@article{rate_inferred_around_GW150914,
   title={THE RATE OF BINARY BLACK HOLE MERGERS INFERRED FROM ADVANCED LIGO OBSERVATIONS SURROUNDING GW150914},
   volume={833},
   ISSN={2041-8213},
   url={http://dx.doi.org/10.3847/2041-8205/833/1/L1},
   DOI={10.3847/2041-8205/833/1/l1},
   number={1},
   journal={Astrophys. J. Lett.},
   publisher={American Astronomical Society},
   author={Abbott, B. P. and others},
   year={2016},
   month=nov, pages={L1}
}

@article{Lynch_2018,
   title={Observational Implications of Lowering the LIGO-Virgo Alert Threshold},
   volume={861},
   ISSN={2041-8213},
   url={http://dx.doi.org/10.3847/2041-8213/aacf9f},
   DOI={10.3847/2041-8213/aacf9f},
   number={2},
   journal={Astrophys. J. Lett.},
   publisher={American Astronomical Society},
   author={Lynch, Ryan and Coughlin, Michael and Vitale, Salvatore and Stubbs, Christopher W. and Katsavounidis, Erik},
   year={2018},
   month=jul, pages={L24}
}

@article{Kapadia_2020,
   title={A self-consistent method to estimate the rate of compact binary coalescences with a Poisson mixture model},
   volume={37},
   ISSN={1361-6382},
   url={http://dx.doi.org/10.1088/1361-6382/ab5f2d},
   DOI={10.1088/1361-6382/ab5f2d},
   number={4},
   journal={Class. Quantum Gravity},
   publisher={IOP Publishing},
   author={Kapadia, Shasvath J and Caudill, Sarah and Creighton, Jolien D E and Farr, Will M and Mendell, Gregory and Weinstein, Alan and Cannon, Kipp and Fong, Heather and Godwin, Patrick and Lo, Rico K L and Magee, Ryan and Meacher, Duncan and Messick, Cody and Mohite, Siddharth R and Mukherjee, Debnandini and Sachdev, Surabhi},
   year={2020},
   month=jan, pages={045007} }

@misc{Ray_2023,
      title={When to Point Your Telescopes: Gravitational Wave Trigger Classification for Real-Time Multi-Messenger Followup Observations}, 
      author={Anarya Ray and others},
      year={2023},
      eprint={2306.07190},
      archivePrefix={arXiv},
      primaryClass={gr-qc},
      url={https://arxiv.org/abs/2306.07190}, 
}

@misc{gwtc-4_pop,
      title={GWTC-4.0: Population Properties of Merging Compact Binaries},
      author={A. G. Abac and others},
      year={2025},
      eprint={2508.18083},
      archivePrefix={arXiv},
      primaryClass={astro-ph.HE},
      url={https://arxiv.org/abs/2508.18083},
}

@article{Mandel_2019,
   title={Extracting distribution parameters from multiple uncertain observations with selection biases},
   volume={486},
   ISSN={1365-2966},
   url={http://dx.doi.org/10.1093/mnras/stz896},
   DOI={10.1093/mnras/stz896},
   number={1},
   journal={Mon. Not. R. Astron. Soc.},
   publisher={Oxford University Press (OUP)},
   author={Mandel, Ilya and Farr, Will M and Gair, Jonathan R},
   year={2019},
   month=mar, pages={1086–1093}
}

@misc{Essick_2025,
      title={Compact Binary Coalescence Sensitivity Estimates with Injection Campaigns during the LIGO-Virgo-KAGRA Collaborations' Fourth Observing Run}, 
      author={Reed Essick and Michael W. Coughlin and Michael Zevin and Deep Chatterjee and Teagan A. Clarke and Storm Colloms and Utkarsh Mali and Simona Miller and Nathan Steinle and Pratyusava Baral and Amanda C. Baylor and Gareth Cabourn Davies and Thomas Dent and Prathamesh Joshi and Praveen Kumar and Cody Messick and Tanmaya Mishra and Amazigh Ouzriat and Khun Sang Phukon and Lorenzo Piccari and Marion Pillas and Max Trevor and Thomas A. Callister and Maya Fishbach},
      year={2025},
      eprint={2508.10638},
      archivePrefix={arXiv},
      primaryClass={gr-qc},
      url={https://arxiv.org/abs/2508.10638}, 
}

@ARTICLE{Schutz_1986,
       author = {{Schutz}, B.~F.},
        title = "{Determining the Hubble constant from gravitational wave observations}",
      journal = {Nature},
         year = 1986,
        month = sep,
       volume = {323},
       number = {6086},
        pages = {310-311},
          doi = {10.1038/323310a0},
       adsurl = {https://ui.adsabs.harvard.edu/abs/1986Natur.323..310S}
}

@article{Petiteau_2011,
   title={CONSTRAINING THE DARK ENERGY EQUATION OF STATE USINGLISAOBSERVATIONS OF SPINNING MASSIVE BLACK HOLE BINARIES},
   volume={732},
   ISSN={1538-4357},
   url={http://dx.doi.org/10.1088/0004-637X/732/2/82},
   DOI={10.1088/0004-637x/732/2/82},
   number={2},
   journal={Astrophys. J.},
   publisher={American Astronomical Society},
   author={Petiteau, Antoine and Babak, Stanislav and Sesana, Alberto},
   year={2011},
   month=apr, 
   pages={82} 
}

@article{Taylor_2012,
  title = {Cosmology using advanced gravitational-wave detectors alone},
  author = {Taylor, Stephen R. and Gair, Jonathan R. and Mandel, Ilya},
  journal = {Phys. Rev. D},
  volume = {85},
  issue = {2},
  pages = {023535},
  numpages = {22},
  year = {2012},
  month = {Jan},
  publisher = {American Physical Society},
  doi = {10.1103/PhysRevD.85.023535},
  url = {https://link.aps.org/doi/10.1103/PhysRevD.85.023535}
}

@article{Pozzo_2012,
  title = {Inference of cosmological parameters from gravitational waves: Applications to second generation interferometers},
  author = {Del Pozzo, Walter},
  journal = {Phys. Rev. D},
  volume = {86},
  issue = {4},
  pages = {043011},
  numpages = {13},
  year = {2012},
  month = {Aug},
  publisher = {American Physical Society},
  doi = {10.1103/PhysRevD.86.043011},
  url = {https://link.aps.org/doi/10.1103/PhysRevD.86.043011}
}

@article{GW170817_first_standard_siren,
   title={A gravitational-wave standard siren measurement of the Hubble constant},
   volume={551},
   ISSN={1476-4687},
   url={http://dx.doi.org/10.1038/nature24471},
   DOI={10.1038/nature24471},
   number={7678},
   journal={Nature},
   publisher={Springer Science and Business Media LLC},
   author={Abbott, B. P. and others},
   year={2017},
   month=oct, 
   pages={85–88} 
}

@article{GW170817_wo_em,
	doi = {10.3847/2041-8213/aaf96e},
	url = {https://doi.org/10.3847/2041-8213/aaf96e},
	year = {2019},
	month = {jan},
	publisher = {The American Astronomical Society},
	volume = {871},
	number = {1},
	pages = {L13},
	author = {Fishbach, M. and others},
	title = {A Standard Siren Measurement of the Hubble Constant from GW170817 without the Electromagnetic Counterpart},
	journal = {Astrophys. J. Lett.}
}

@article{Soares_Santos_2019,
   title={First Measurement of the Hubble Constant from a Dark Standard Siren using the Dark Energy Survey Galaxies and the LIGO/Virgo Binary–Black-hole Merger GW170814},
   volume={876},
   ISSN={2041-8213},
   url={http://dx.doi.org/10.3847/2041-8213/ab14f1},
   DOI={10.3847/2041-8213/ab14f1},
   number={1},
   journal={Astrophys. J. Lett.},
   publisher={American Astronomical Society},
   author={Soares-Santos, M. and others},
   year={2019},
   month=apr, pages={L7} 
}

@article{Gray_2020,
   title={Cosmological inference using gravitational wave standard sirens: A mock data analysis},
   volume={101},
   ISSN={2470-0029},
   url={http://dx.doi.org/10.1103/PhysRevD.101.122001},
   DOI={10.1103/physrevd.101.122001},
   number={12},
   journal={Phys. Rev. D},
   publisher={American Physical Society (APS)},
   author={Gray, Rachel and Hernandez, Ignacio Magaña and Qi, Hong and Sur, Ankan and Brady, Patrick R. and Chen, Hsin-Yu and Farr, Will M. and Fishbach, Maya and Gair, Jonathan R. and Ghosh, Archisman and Holz, Daniel E. and Mastrogiovanni, Simone and Messenger, Christopher and Steer, Danièle A. and Veitch, John},
   year={2020},
   month=jun }

@article{Farr_2019,
   title={A Future Percent-level Measurement of the Hubble Expansion at Redshift 0.8 with Advanced LIGO},
   volume={883},
   ISSN={2041-8213},
   url={http://dx.doi.org/10.3847/2041-8213/ab4284},
   DOI={10.3847/2041-8213/ab4284},
   number={2},
   journal={Astrophys. J. Lett.},
   publisher={American Astronomical Society},
   author={Farr, Will M. and Fishbach, Maya and Ye, Jiani and Holz, Daniel E.},
   year={2019},
   month=oct, pages={L42}
}

@article{Mapelli_2020,
   title={Binary Black Hole Mergers: Formation and Populations},
   volume={7},
   ISSN={2296-987X},
   url={http://dx.doi.org/10.3389/fspas.2020.00038},
   DOI={10.3389/fspas.2020.00038},
   journal={Front. Astron. Space Sci.},
   publisher={Frontiers Media SA},
   author={Mapelli, Michela},
   year={2020},
   month=jul
}

@article{Ezquiaga_2022,
  title = {Spectral Sirens: Cosmology from the Full Mass Distribution of Compact Binaries},
  author = {Ezquiaga, Jose Mar\'{\i}a and Holz, Daniel E.},
  journal = {Phys. Rev. Lett.},
  volume = {129},
  issue = {6},
  pages = {061102},
  numpages = {6},
  year = {2022},
  month = {Aug},
  publisher = {American Physical Society},
  doi = {10.1103/PhysRevLett.129.061102},
  url = {https://link.aps.org/doi/10.1103/PhysRevLett.129.061102}
}

@article{Mukherjee_2022,
   title={The redshift dependence of black hole mass distribution: is it reliable for standard sirens cosmology?},
   volume={515},
   ISSN={1365-2966},
   url={http://dx.doi.org/10.1093/mnras/stac2152},
   DOI={10.1093/mnras/stac2152},
   number={4},
   journal={Mon. Not. R. Astron. Soc.},
   publisher={Oxford University Press (OUP)},
   author={Mukherjee, Suvodip},
   year={2022},
   month=aug, 
   pages={5495–5505} 
}

@article{Yu_2022,
    author = "Yu, Hang and Seymour, Brian and Wang, Yijun and Chen, Yanbei",
    title = "{Uncertainty and Bias of Cosmology and Astrophysical Population Model from Statistical Dark Sirens}",
    eprint = "2206.09984",
    archivePrefix = "arXiv",
    primaryClass = "astro-ph.CO",
    doi = "10.3847/1538-4357/ac9da0",
    journal = "Astrophys. J.",
    volume = "941",
    number = "2",
    pages = "174",
    year = "2022"
}

@misc{icarogw,
      title={ICAROGW: A python package for inference of astrophysical population properties of noisy, heterogeneous and incomplete observations},
      author={Simone Mastrogiovanni and Grégoire Pierra and Stéphane Perriès and Danny Laghi and Giada Caneva Santoro and Archisman Ghosh and Rachel Gray and Christos Karathanasis and Konstantin Leyde},
      year={2023},
      eprint={2305.17973},
      archivePrefix={arXiv},
      primaryClass={astro-ph.CO},
      url={https://arxiv.org/abs/2305.17973},
}

@article{gwtc-3_cosmo,
doi = {10.3847/1538-4357/ac74bb},
url = {https://dx.doi.org/10.3847/1538-4357/ac74bb},
year = {2023},
month = {jun},
publisher = {The American Astronomical Society},
volume = {949},
number = {2},
pages = {76},
author = {R. Abbott and others},
title = {Constraints on the Cosmic Expansion History from GWTC–3},
journal = {Astrophys. J.}
}

@article{Gray_2023,
doi = {10.1088/1475-7516/2023/12/023},
url = {https://doi.org/10.1088/1475-7516/2023/12/023},
year = {2023},
month = {dec},
publisher = {IOP Publishing},
volume = {2023},
number = {12},
pages = {023},
author = {Gray, Rachel and Beirnaert, Freija and Karathanasis, Christos and Revenu, Benoît and Turski, Cezary and Chen, Anson and Baker, Tessa and Vallejo, Sergio and Romano, Antonio Enea and Ghosh, Tathagata and Ghosh, Archisman and Leyde, Konstantin and Mastrogiovanni, Simone and More, Surhud},
title = {Joint cosmological and gravitational-wave population inference using dark sirens and galaxy catalogues},
journal = {J. Cosmol. Astropart. Phys.}
}

@misc{gwtc-4_cosmo,
      title={GWTC-4.0: Constraints on the Cosmic Expansion Rate and Modified Gravitational-wave Propagation}, 
      author={The LIGO Scientific Collaboration and the Virgo Collaboration and the KAGRA Collaboration},
      year={2025},
      eprint={2509.04348},
      archivePrefix={arXiv},
      primaryClass={astro-ph.CO},
      url={https://arxiv.org/abs/2509.04348}, 
}

@article{Beirnaert_2025,
   title={A Hubble constant estimation with dark standard sirens and galaxy cluster catalogues},
   volume={542},
   ISSN={1365-2966},
   url={http://dx.doi.org/10.1093/mnras/staf1432},
   DOI={10.1093/mnras/staf1432},
   number={4},
   journal={Mon. Not. R. Astron. Soc.},
   publisher={Oxford University Press (OUP)},
   author={Beirnaert, Freija and Dálya, Gergely and Ghosh, Archisman},
   year={2025},
   month=aug, pages={3346–3353}
}

@misc{Mali_2024,
      title={Striking a Chord with Spectral Sirens: multiple features in the compact binary population correlate with $H_0$}, 
      author={Utkarsh Mali and Reed Essick},
      year={2024},
      eprint={2410.07416},
      archivePrefix={arXiv},
      primaryClass={astro-ph.HE},
      url={https://arxiv.org/abs/2410.07416}, 
}

@article{Cannon_2013,
  title = {Method to estimate the significance of coincident gravitational-wave observations from compact binary coalescence},
  author = {Cannon, Kipp and Hanna, Chad and Keppel, Drew},
  journal = {Phys. Rev. D},
  volume = {88},
  issue = {2},
  pages = {024025},
  numpages = {9},
  year = {2013},
  month = {Jul},
  publisher = {American Physical Society},
  doi = {10.1103/PhysRevD.88.024025},
  url = {https://link.aps.org/doi/10.1103/PhysRevD.88.024025}
}

@misc{cannon2015likelihoodratiorankingstatisticcompact,
      title={Likelihood-Ratio Ranking Statistic for Compact Binary Coalescence Candidates with Rate Estimation}, 
      author={Kipp Cannon and Chad Hanna and Jacob Peoples},
      year={2015},
      eprint={1504.04632},
      archivePrefix={arXiv},
      primaryClass={astro-ph.IM},
      url={https://arxiv.org/abs/1504.04632}, 
}

@article{Messick_2017,
   title={Analysis framework for the prompt discovery of compact binary mergers in gravitational-wave data},
   volume={95},
   ISSN={2470-0029},
   url={http://dx.doi.org/10.1103/PhysRevD.95.042001},
   DOI={10.1103/physrevd.95.042001},
   number={4},
   journal={Phys. Rev. D},
   publisher={American Physical Society (APS)},
   author={Messick, Cody and others},
   year={2017},
   month=feb
}

@phdthesis{Fong_thesis,
	author = {Fong, Heather Kin Yee},
	title = {From simulations to signals: Analyzing gravitational waves from compact binary coalescences},
	school = {University of Toronto},
	year = {2018}
}

@article{Tsukada_2023,
  title = {Improved ranking statistics of the GstLAL inspiral search for compact binary coalescences},
  author = {Tsukada, Leo and others},
  journal = {Phys. Rev. D},
  volume = {108},
  issue = {4},
  pages = {043004},
  numpages = {13},
  year = {2023},
  month = {Aug},
  publisher = {American Physical Society},
  doi = {10.1103/PhysRevD.108.043004},
  url = {https://link.aps.org/doi/10.1103/PhysRevD.108.043004}
}

@article{Ewing_2024,
  title = {Performance of the low-latency GstLAL inspiral search towards LIGO, Virgo, and KAGRA's fourth observing run},
  author = {Ewing, Becca and Huxford, Rachael and Singh, Divya and Tsukada, Leo and Hanna, Chad and Huang, Yun-Jing and Joshi, Prathamesh and Li, Alvin K. Y. and Magee, Ryan and Messick, Cody and Pace, Alex and Ray, Anarya and Sachdev, Surabhi and Sakon, Shio and Tapia, Ron and Adhicary, Shomik and Baral, Pratyusava and Baylor, Amanda and Cannon, Kipp and Caudill, Sarah and Chaudhary, Sushant Sharma and Coughlin, Michael W. and Cousins, Bryce and Creighton, Jolien D. E. and Essick, Reed and Fong, Heather and George, Richard N. and Godwin, Patrick and Harada, Reiko and Kennington, James and Kuwahara, Soichiro and Meacher, Duncan and Morisaki, Soichiro and Mukherjee, Debnandini and Niu, Wanting and Posnansky, Cort and Toivonen, Andrew and Tsutsui, Takuya and Ueno, Koh and Viets, Aaron and Wade, Leslie and Wade, Madeline and Waratkar, Gaurav},
  journal = {Phys. Rev. D},
  volume = {109},
  issue = {4},
  pages = {042008},
  numpages = {18},
  year = {2024},
  month = {Feb},
  publisher = {American Physical Society},
  doi = {10.1103/PhysRevD.109.042008},
  url = {https://link.aps.org/doi/10.1103/PhysRevD.109.042008}
}

@misc{gstlal,
	title = {{GstLAL}}, 
	howpublished = {\url{https://git.ligo.org/lscsoft/gstlal}},
	note = {Accessed: 2025-09-11}
}

@misc{s5s6_2021_reanalysis,
  author       = {Heather Fong and others},
  title        = {{GW070605: An Undisclosed Binary Neutron Star Hardware Injection in LIGO’s Fifth Science Run}},
  note         = {in preparation}
}

@misc{LIGO-T1800044,
	author = {L. Barsotti and S. Gras and M. Evans and P. Fritschel},

	title = {{U}pdated {A}dvanced {LIGO} sensitivity design curve},
	  
	year = {2018},

	howpublished = {\url{https://dcc.ligo.org/LIGO-T1800044/public}}
}

@misc{LIGO-P1600143,
	author = {B. P. Abbott and others},

	title = {{E}xploring the {S}ensitivity of {N}ext {G}eneration {G}ravitational {W}ave {D}etectors},
	  
	year = {2016},

	howpublished = {\url{https://dcc.ligo.org/LIGO-P1600143/public}}
}

@inbook{Creighton_2011,
author = {Creighton, Jolien D. E. and Anderson, Warren G.},
publisher = {John Wiley \& Sons, Ltd},
title = {Gravitational-Wave Data Analysis},
booktitle = {Gravitational‐Wave Physics and Astronomy},
year = {2011}
}

@article{GW150914,
  title = {Observation of Gravitational Waves from a Binary Black Hole Merger},
  author = {Abbott, B. P. and others},
  collaboration = {LIGO Scientific Collaboration and Virgo Collaboration},
  journal = {Phys. Rev. Lett.},
  volume = {116},
  issue = {6},
  pages = {061102},
  numpages = {16},
  year = {2016},
  month = {Feb},
  publisher = {American Physical Society},
  doi = {10.1103/PhysRevLett.116.061102},
  url = {https://link.aps.org/doi/10.1103/PhysRevLett.116.061102}
}

@article{GWTC-1,
  title = {GWTC-1: A Gravitational-Wave Transient Catalog of Compact Binary Mergers Observed by LIGO and Virgo during the First and Second Observing Runs},
  author = {Abbott, B. P. and others},
  collaboration = {LIGO Scientific Collaboration and Virgo Collaboration},
  journal = {Phys. Rev. X},
  volume = {9},
  issue = {3},
  pages = {031040},
  numpages = {49},
  year = {2019},
  month = {Sep},
  publisher = {American Physical Society},
  doi = {10.1103/PhysRevX.9.031040},
  url = {https://link.aps.org/doi/10.1103/PhysRevX.9.031040}
}

@article{GWTC-2,
  title = {GWTC-2: Compact Binary Coalescences Observed by LIGO and Virgo during the First Half of the Third Observing Run},
  author = {Abbott, R. and others},
  collaboration = {LIGO Scientific Collaboration and Virgo Collaboration},
  journal = {Phys. Rev. X},
  volume = {11},
  issue = {2},
  pages = {021053},
  numpages = {52},
  year = {2021},
  month = {Jun},
  publisher = {American Physical Society},
  doi = {10.1103/PhysRevX.11.021053},
  url = {https://link.aps.org/doi/10.1103/PhysRevX.11.021053}
}

@article{GWTC-2.1,
  title = {GWTC-2.1: Deep extended catalog of compact binary coalescences observed by LIGO and Virgo during the first half of the third observing run},
  author = {Abbott, R. and others},
  collaboration = {The LIGO Scientific Collaboration and the Virgo Collaboration},
  journal = {Phys. Rev. D},
  volume = {109},
  issue = {2},
  pages = {022001},
  numpages = {45},
  year = {2024},
  month = {Jan},
  publisher = {American Physical Society},
  doi = {10.1103/PhysRevD.109.022001},
  url = {https://link.aps.org/doi/10.1103/PhysRevD.109.022001}
}

@article{GWTC-3,
  title = {GWTC-3: Compact Binary Coalescences Observed by LIGO and Virgo during the Second Part of the Third Observing Run},
  author = {Abbott, R. and others},
  collaboration = {LIGO Scientific Collaboration, Virgo Collaboration, and KAGRA Collaboration},
  journal = {Phys. Rev. X},
  volume = {13},
  issue = {4},
  pages = {041039},
  numpages = {89},
  year = {2023},
  month = {Dec},
  publisher = {American Physical Society},
  doi = {10.1103/PhysRevX.13.041039},
  url = {https://link.aps.org/doi/10.1103/PhysRevX.13.041039}
}

@misc{GWTC-4.0,
      title={{GWTC-4.0: Updating the Gravitational-Wave Transient Catalog with Observations from the First Part of the Fourth LIGO-Virgo-KAGRA Observing Run}},
      author={A. G. Abac and others},
      year={2025},
      eprint={2508.18082},
      archivePrefix={arXiv},
      primaryClass={gr-qc},
      url={https://arxiv.org/abs/2508.18082},
}

@article{Gabbard_2021,
   title={Bayesian parameter estimation using conditional variational autoencoders for gravitational-wave astronomy},
   volume={18},
   ISSN={1745-2481},
   url={http://dx.doi.org/10.1038/s41567-021-01425-7},
   DOI={10.1038/s41567-021-01425-7},
   number={1},
   journal={Nat. Phys.},
   publisher={Springer Science and Business Media LLC},
   author={Gabbard, Hunter and Messenger, Chris and Heng, Ik Siong and Tonolini, Francesco and Murray-Smith, Roderick},
   year={2021},
   month=dec, 
   pages={112–117} 
}

@article{DINGO,
  title = {Real-Time Gravitational Wave Science with Neural Posterior Estimation},
  author = {Dax, Maximilian and Green, Stephen R. and Gair, Jonathan and Macke, Jakob H. and Buonanno, Alessandra and Sch\"olkopf, Bernhard},
  journal = {Phys. Rev. Lett.},
  volume = {127},
  issue = {24},
  pages = {241103},
  numpages = {7},
  year = {2021},
  month = {Dec},
  publisher = {American Physical Society},
  doi = {10.1103/PhysRevLett.127.241103},
  url = {https://link.aps.org/doi/10.1103/PhysRevLett.127.241103}
}

@article{Bhardwaj_2023,
  title = {Sequential simulation-based inference for gravitational wave signals},
  author = {Bhardwaj, Uddipta and Alvey, James and Miller, Benjamin Kurt and Nissanke, Samaya and Weniger, Christoph},
  journal = {Phys. Rev. D},
  volume = {108},
  issue = {4},
  pages = {042004},
  numpages = {21},
  year = {2023},
  month = {Aug},
  publisher = {American Physical Society},
  doi = {10.1103/PhysRevD.108.042004},
  url = {https://link.aps.org/doi/10.1103/PhysRevD.108.042004}
}

@misc{AMPLFI,
      title={Rapid Likelihood Free Inference of Compact Binary Coalescences using Accelerated Hardware},
      author={Deep Chatterjee and Ethan Marx and William Benoit and Ravi Kumar and Malina Desai and Ekaterina Govorkova and Alec Gunny and Eric Moreno and Rafia Omer and Ryan Raikman and Muhammed Saleem and Shrey Aggarwal and Michael W. Coughlin and Philip Harris and Erik Katsavounidis},
      year={2024},
      eprint={2407.19048},
      archivePrefix={arXiv},
      primaryClass={gr-qc},
      url={https://arxiv.org/abs/2407.19048},
}

@misc{Aframe,
      title={A machine-learning pipeline for real-time detection of gravitational waves from compact binary coalescences},
      author={Ethan Marx and William Benoit and Alec Gunny and Rafia Omer and Deep Chatterjee and Ricco C. Venterea and Lauren Wills and Muhammed Saleem and Eric Moreno and Ryan Raikman and Ekaterina Govorkova and Dylan Rankin and Michael W. Coughlin and Philip Harris and Erik Katsavounidis},
      year={2025},
      eprint={2403.18661},
      archivePrefix={arXiv},
      primaryClass={gr-qc},
      url={https://arxiv.org/abs/2403.18661},
}

@misc{lalsuite,
       author         = "{LIGO Scientific Collaboration}",
       title          = "{LIGO} {A}lgorithm {L}ibrary - {LALS}uite",
       howpublished   = "free software (GPL)",
       doi            = "10.7935/GT1W-FZ16",
       year           = "2018"
 }
\end{document}